# FRIEDRICH-ALEXANDER-UNIVERSITÄT ERLANGEN-NÜRNBERG

# Pay Clauses in Public Procurement: The Wage Impact of Collective Bargaining Compliance Laws in Germany

Vinzenz Pyka

Juni 2025

# Pay Clauses in Public Procurement: The Wage Impact of Collective Bargaining Compliance Laws in Germany[*]

Vinzenz Pyka[†]

**Abstract:** Using administrative data from Germany, this study provides first evidence on the wage effects of collective bargaining compliance laws. These laws require establishments receiving public contracts to pay wages set by a representative collective agreement, even if they are not formally bound by one. Leveraging variation in the timing of law implementation across federal states, and focusing on the public transport sector — where regulation is uniform and demand is driven solely by state-level needs — I estimate dynamic treatment effects using event-study designs. The results indicate that within five years of the law's implementation, wage increases were on average 2.9 to 4.6 per cent higher in federal states with such a law compared to those without one — but only in East Germany. These findings highlight the potential for securing collectively agreed wages in times of declining collective bargaining coverage.

**Zusammenfassung:** Ich nutze administrative Daten aus Deutschland und liefere erstmals empirische Evidenz zu den Lohneffekten von Tariftreuegesetzen. Diese Gesetze verpflichten Unternehmen, die öffentliche Aufträge erhalten, ihre Beschäftigten nach einem repräsentativen Tarifvertrag zu entlohnen — auch wenn sie selbst nicht tarifgebunden sind. Auf Grundlage der Gesetzeseinführungen in den Bundesländern analysiere ich mithilfe von Event-Study-Modellen die Lohnentwicklung von Beschäftigten im öffentlichen Nahverkehr. Dieser Sektor eignet sich besonders gut zur Evaluation, da er bundesweit einheitlich reguliert ist und die Nachfrage ausschließlich durch den Staat bestimmt wird. Die Ergebnisse zeigen, dass in Bundesländern mit einem Tariftreuegesetz die Löhne ceteris paribus im Durchschnitt innerhalb von fünf Jahren nach der Einführung um 2,9 bis 4,6 Prozent stärker gestiegen sind als in Bundesländern ohne Tariftreuegesetz — allerdings nur in Ostdeutschland. Dies legt nahe, dass ein Bundestariftreuegesetz beitragen könnte, Tariflöhne angesichts einer sinkenden Tarifbindung zu sichern.

[*] The author thanks Claus Schnabel, Georg Caspers, Hannes Walz, Stephanie Prümer, Matthias Collischon, André Rieder, Benedikt Schröpf, Jens Stegmaier, and Jan Weikl for valuable feedback, Maximilian Aurbach for insights into public transport, and Vanessa Bartsch for excellent research assistance.
[†] Friedrich-Alexander-Universität Erlangen-Nürnberg, Nürnberg, Germany; correspondence to: Vinzenz Pyka, University of Erlangen-Nürnberg, School of Business, Economics and Society, Lange Gasse 20, 90403 Nürnberg, Germany. E-Mail: vinzenz.pyka@fau.de

# 1. Introduction

Over[1] the past few decades, labour market research has increasingly focused on examining the wage and employment effects of state interventions, such as minimum wages (e.g. Bossler & Gerner, 2019; Bossler & Schank, 2023; Burauel et al., 2020; Card & Krueger, 2000; Dustmann et al., 2021). Simultaneously, empirical studies in Germany have spotlighted the growing interest in union wage premiums and the impact of collective bargaining agreements on wages (Addison et al., 2014; Bonaccolto-Töpfer & Schnabel, 2023; Hirsch & Mueller, 2020; Jäger et al., 2024). Within this context, public procurement has emerged in political discourse as a crucial tool for labour market regulation against the backdrop of declining collective bargaining coverage in Germany (Sack & Sarter, 2018; Schulten, 2021). In 2000, 70 per cent of employees in West Germany and 56 per cent in East Germany were covered by collective agreements, but by 2022, this had fallen to only 52 per cent in West Germany and 45 per cent in East Germany (Hohendanner & Kohaut, 2024).

In 2025, Germany's governing parties committed in their coalition agreement to introduce a collective bargaining compliance law (*Tariftreuegesetz*) to expand collective bargaining coverage and establish collectively negotiated wages as the standard in the labour market (CDU et al., 2025).[2] Such a law requires companies receiving public contracts to ensure that their employees receive working conditions and pay rates according to a representative collective agreement, even if the companies themselves are not bound by a collective agreement with a trade union. This requirement in the awarding of public contracts aims to offset the potential wage cost disadvantages faced by companies with collective agreements when competing for public contracts against those not bound by collective agreements. However, the economic impact of this regulation remains uncertain. Companies may circumvent the wage floor by opting for private contracts that are free from these restrictions. Conversely, wages might increase if companies accept government contracts and adhere to wage conditions. Moreover, referring to existing collective agreements reduces information barriers for workers in non-unionised companies, potentially pressuring these companies to raise wages to retain employees and align with industry standards. Overall, the wage effect remains ambiguous.

---

[1] Artificial intelligence, like ChatGPT and DeepL, was utilised to enhance the text by refining its grammar and sentence structure.

[2] The previous federal government, composed of the Social Democrats (SPD), the Greens (Bündnis 90/Die Grünen), and the Liberals (FDP), had already drafted a bill for a collective bargaining compliance law in 2024. However, after the coalition broke apart in 2024, the draft was never brought to a vote in parliament (Bundesministerium für Arbeit und Soziales, 2024).

Internationally, a collective bargaining compliance law is comparable to living wage policies in the US and UK (Datta & Machin, 2024; Johnson, 2017; Neumark et al., 2012). According to Datta and Machin (2024), living wages are typically calculated based on a consumption bundle defined to reach a minimum standard of living, and they are higher than the mandated minimum wage. Firms with procurement contracts from municipal governments are required to pay workers the living wage. The idea behind the German law is similar, but it differs in how the procurement-specific wage is set. In Germany, the wage is not defined based on a consumption bundle. Instead, the law uses wages from existing collective agreements and refers to these agreements as representative collective agreements for specific public procurement contracts. Nevertheless, the underlying idea of these laws is comparable, as they establish the use of pay clauses in public procurement.

This study empirically analyses the wage impact of collective bargaining compliance laws by exploiting variation in their introduction at the federal state level in Germany. To the best of my knowledge, this is the first study to examine wage effects by leveraging the implementation of collective bargaining compliance laws in Germany, using administrative data. Therefore, the study is contributing to the existing literature on collective bargaining premia (Addison et al., 2014; Bonaccolto-Töpfer & Schnabel, 2023; Hirsch & Mueller, 2020; Jäger et al., 2024), minimum wages (Bossler & Gerner, 2019; Bossler & Schank, 2023; Burauel et al., 2020; Card & Krueger, 2000; Dustmann et al., 2021) and living wages (Datta & Machin, 2024; Johnson, 2017; Neumark et al., 2012). It quantifies the effects of collective bargaining compliance laws, using administrative data and staggered event study methods to assess their impact on public transport employees' wages. This study contributes to the debate on labour regulations by providing initial insights into how wages might increase if the state mandates collectively agreed wages for public contracts, offering valuable evidence on how collectively agreed wages can be secured in the context of declining collective bargaining coverage.

The main findings, drawn from dynamic two-way fixed effects (TWFE) and Callaway and Sant'Anna (2021) event studies, reveal that wages for public transport employees increased by 2.9 to 4.6 per cent within five years of the law's implementation - an effect observed exclusively in East Germany. In contrast, no wage effects were detected in West Germany based on Callaway and Sant'Anna (2021) estimations, including establishment fixed effects. My main findings hold up under various robustness checks, including considering the impact of introducing the federal minimum wage in Germany in 2015.

The structure of the study is as follows: Section two provides a historical overview of collective bargaining compliance laws in Germany. Following this, section three examines the specific characteristics of the public transport sector in relation to these laws and public procurement practices. The data and empirical strategy are described in section four, while descriptive insights are presented in section five. Section six outlines the empirical model. In section seven, the main results are reported and discussed, and robustness checks are included. Finally, section eight concludes with a summary and policy implications.

**2. Collective bargaining compliance laws in Germany**

In Germany, public contracts are generally awarded with the aim of using budgetary funds efficiently and economically, which typically leads to contracts being granted to the lowest bidder. This practice raises concerns that companies might be incentivised to avoid joining or to withdraw from collective bargaining agreements, as such agreements could put them at a cost disadvantage compared to competitors who are not bound by these agreements. To address this disincentive, trade unions proposed a collective bargaining compliance law that mandates companies receiving public contracts to pay the wages and provide the working conditions outlined in a specified representative collective agreement to their employees, even if the company itself is not bound by such an agreement. While trade unions in Germany advocate for collective bargaining compliance laws, employer representatives, on the other hand, view these laws as an additional bureaucratic burden (Bundesministerium für Arbeit und Soziales, 2023).

The current debate on a federal-level law builds on a long-standing evolution in state-level collective bargaining regulations that has been ongoing since the early 2000s (Schulten, 2021). Initially, these laws were introduced in response to a decline in collective bargaining coverage and fears that the free movement of labour—associated with the EU's eastward enlargement in the 2000s—would lead to a deterioration in wages and working conditions in Germany. The first state-level laws initially applied to the construction industry and public transport but were gradually extended to other sectors (Schulten, 2021).

In the mid-2000s, questions arose regarding the legality of such state laws. In principle, collective agreements on wages and working conditions are typically negotiated and set by trade unions and employers' associations without state intervention in Germany. This right is legally protected by the German constitution (Oberfichtner & Schnabel, 2019). Therefore, any state intervention in wage setting raises the question of its compatibility with the constitution.

In 2006, the Federal Constitutional Court ruled that the collective bargaining compliance laws and associated pay regulations were consistent with the constitution (Bundesverfassungsgericht, 2006). However, this precedent was overturned by the European Court of Justice in the 2008 *Rüffert* ruling, which concerned Lower Saxony's collective bargaining compliance laws. The Court found that the reference to a representative collective agreement, instead of a generally binding agreement, was incompatible with European law. The Court criticised the fact that collective bargaining compliance requirements applied only to public contracts, questioning why work done under a public contract should be treated differently from work under a private contract. This ruling led to the suspension and subsequent revision of laws in federal states, which then referred to generally binding collective agreements (Caspers, 2024; McCrudden, 2011; Schulten, 2021; Thüsing, 2022).

Despite this, the European Court of Justice's main criticism was somewhat softened after 2008. Before the introduction of a generally binding minimum wage in Germany, the Court deviated from its earlier position in the 2015 RegioPost judgment (C-115/14) and ruled that procurement-specific minimum wages could be compatible with European law (Schulten, 2021). Consequently, some federal states introduced procurement-specific minimum wages alongside the existing references to collective bargaining agreements in their collective bargaining compliance laws.[3] Moreover, in 2018, the European Union revised the European Posting of Workers Directive, expanding the scope for establishing social standards within procurement laws (Caspers, 2024; Nassibi et al., 2016; Sack & Sarter, 2018; Schulten, 2021).

However, it remains unclear in legal scholarship whether these developments have altered the validity of the Rüffert judgment and whether they allow for references to representative collective agreements — as argued by Schulten (2021) - or not, as argued by Caspers (2024). It remains an open question how the European Court of Justice would rule in a new case regarding the legality of such federal state laws.

---

[3] For an overview of which federal states have introduced procurement-specific minimum wages, see Sack and Sarter (2018).

## 3. Regulations for the public transport sector

### 3.1 Evolution of collective bargaining compliance laws in the public transport sector

Public transport is considered part of the provision of services of general interest and is clearly defined in both federal and state legislation in Germany.[4] It is organised by municipalities, independent cities, or federal states[5] and holds a special status under European law.[6] This has led to the legal interpretation that the European Court of Justice's 2008 Rüffert ruling does not apply to local public transport (for a more detailed legal analysis, see Thüsing (2022)). The Rüffert judgment prompted changes in federal state legislation, and while generally binding collective agreements are required in other sectors, references to representative collective agreements in the public transport are still included in the laws (Schulten, 2021; Thüsing, 2022). This legal distinction is crucial to my identification strategy and explains why the focus of my analysis is on local public transport. It ensures that the legal framework for public transport has remained consistent over time and is comparable across federal states, allowing for cross-state comparisons.

(Table 1 about here)

Table 1 provides an overview of the implementation of collective bargaining compliance laws concerning public transport across the federal states, divided into West and East Germany. With the exception of Berlin, all East German federal states adopted these laws following the 2008 Rüffert ruling. In contrast, the situation in West

---

[4] The Passenger Transport Act (PBefG) §8 defines local public passenger transport as the generally accessible transport of passengers by trams, buses, and motor vehicles on scheduled services, primarily intended to meet the demand for transport in urban, suburban, or regional areas. In case of doubt, this applies if the total distance travelled by the means of transport does not exceed 50 kilometres, or if the total journey time does not exceed one hour.

[5] In Germany, local public transport is managed by the federal states and consists of general local transport and local rail transport. Most federal state laws assign contract awarding for general local transport to administrative districts or independent cities, while local rail transport contracts are awarded by the state or these local entities.

[6] See Articles 90 to 100 on the Treaty on the Functioning of the European Union (European Union, 2016).

Germany is more varied, with laws being implemented both before and after 2008. When introduced, most federal states set the threshold for applicability between EUR 10,000 and EUR 50,000. In East Germany, the range of thresholds is more heterogenous; for instance, Mecklenburg-Western Pomerania set no threshold at all, Brandenburg implemented a very low one, while Thuringia and Saxony-Anhalt set thresholds comparable to those in the western states. However, beyond the threshold, all laws — explicitly after 2008 - refer to a representative collective bargaining agreement when awarding public contracts, and are therefore comparable.

### 3.2 Regulations for contract awarding in public transport

A central question within the institutional framework concerns how public transport contracts are awarded. The allocation of public passenger transport services by rail and road is regulated at the European level by Regulation (EC) 1370/2007, which has been in force since 2009 (European Union, 2007). This regulation specifies that the maximum duration of a public contract is ten years for bus services and fifteen years for passenger rail services, while also detailing the procedures for contract awards (for a comprehensive description, see Resch (2015)).

Regulation (EC) 1370/2007 generally mandates that public transport services be awarded through competitive tendering, meaning contracts are publicly tendered and providers compete for them. Such contract awards fall under the collective bargaining compliance laws applicable in federal states. However, the EU regulation permits exceptions where public transport services can be awarded directly, or through in-house awarding, without a public tender. In these cases, collective bargaining compliance laws do not apply, as they are only relevant to public tenders.

The conditions under which contracts can be awarded directly are strictly regulated by European law under Regulation (EC) 1370/2007 (European Union, 2007). Direct awards are permissible if: first, the local authority exerts influence and control over the contracted company, which does not participate in tenders beyond its jurisdiction; second, the contract volume is below a set threshold, with an annual average value under EUR 1 million or fewer than 300,000 km annually in public passenger transport services, or doubled for operators with no more than 23 vehicles; third, it serves as an emergency measure if there is a disruption in service or an imminent risk of such a disruption.[7]

---

[7] For a more detailed comparison between direct awarding and tendering, see Grüttner et al. (2012).

As Resch (2015) discusses, direct awarding, founded on authority control, aligns with German municipal constitutional principles governing public enterprises. Such companies are restricted to their municipal roles and cannot compete externally (Resch, 2015). Brandt and Schulten (2008) emphasize that unlike private local public transport companies, municipal firms tend to have more comprehensive collective bargaining coverage. These authors highlight a dual labour regime in bargaining coverage between public and privatized companies, accentuated after public service privatizations in Germany, where public transport usually maintains high collective agreement coverage and union organization, whereas outsourced or private companies may not (Brandt & Schulten, 2008).

Resch's (2015) and Brandt and Schulten's (2008) findings imply that municipalities might choose the direct award exception when public contracts are awarded to municipally- or state-controlled companies. These companies usually have higher collective bargaining coverage than private ones. Conversely, the tendering procedure, to which collective bargaining compliance laws apply, is likely to be chosen by municipalities when contracts are awarded to private companies. These companies typically have lower collective bargaining coverage than public companies.

The extent of direct awarding usage in Germany remains unclear due to a lack of comprehensive data systematically documenting such contracts. However, the European Court of Auditors (2023) evaluated the implementation of the EU Directive 2014/24/EU, which outlines the general rules for public contract awards across Europe, analysing the period from 2011 to 2021. Contracts reported to the Tender Electronic Daily (TED) system were examined by the European Court of Auditors, with findings indicating that the share of contracts awarded directly in Germany — without tender publications - dropped from 34.7 per cent in 2011 to 18.8 per cent in 2021.[8] Specifically, data related to "Transport services (excl. Waste transport)" indicate that direct awards increased slightly from around 14.0 per cent in 2011 to 16.1 per cent in 2021.

In consideration of the potentially broad categorization by the European Court of Auditors, I restricted my analysis of Tenders Electronic Daily (TED) data to observations where a link to the call for tenders published on TED was available for public transport (road) in the same period, without EU threshold restrictions (Table 2).[9] The analysis revealed that 1.46 per cent of tenders referencing a TED link mentioned direct awards in their title and/or description, whereas over 98 per cent did not mention direct awarding.

---

[8] Tender Electronic Daily (TED) contains all public tenders and contract award notices exceeding EU thresholds, as these must be published on the TED website. Authorities may publish contracts below thresholds in this system, but they were excluded from the European Court of Auditors' analysis. EU thresholds: 5,382,000 EUR for public works; 140,000 EUR for central government contracts; 215,000 EUR for local and regional government contracts; 750,000 EUR for social and other specific service contracts (European Court of Auditors, 2023).

(Table 2 around here)

These descriptions should be interpreted with caution, as the data offer only a partial perspective on contract awarding in Germany, particularly as most sub-threshold contracts are not captured. Nevertheless, they indicate that while direct awarding is used, it does not appear to be the main method employed by public authorities. Concerning the research question on the impact of collective bargaining compliance laws on wages, direct awarding does not seem to be a primary issue. For a sound identification strategy, it is crucial to ensure there are no unobserved, time-varying systematic differences in its application between treated and control federal states. Given the uniform legal framework across municipalities, there seems to be little reason to expect systematic differences in direct awarding preferences on average, between municipalities with and without collective bargaining agreements. Therefore, any unobserved differences in the preference for direct awarding over public tendering should be addressed as effectively as possible by employing difference-in-differences comparisons across the states.

## 4. Data and empirical strategy

This analysis is based on administrative data from the Integrated Labour Market Biographies (SIAB) sample (SIAB regional file 7521), provided by the Institute for Employment Research (IAB) of the Federal Employment Agency, covering the years 2000 to 2020. The SIAB is a two per cent random sample from the Integrated Employment Biographies (IEB) and allows the tracking of individuals' employment histories on a daily basis (Schmucker et al., 2023). This data is available as spell data and has been prepared into annual panel data as described by Stüber et al. (2023).

---

[9] Data originates from opentender.eu (https://opentender.eu/de). Based on the Common Procurement Vocabulary (CPV) system for standardised classifications in public procurement the category 60112000 (Public Transport – Road) is used.

The SIAB provides information on age, gender, education, working status, daily wages, and the federal state of employment. Due to social security contribution limits, wages are imputed according to the method described by Stüber et al. (2023) and deflated to the year 2015. Subsequently, I log-transform this daily wage, which serves as the dependent variable in this study. Furthermore, the SIAB can be linked to the Establishment History Panel (BHP), which provides data on establishment size, the number of full-time and part-time employees, qualification levels, classifications of economic activity, and the county in which the company is located (Schmucker et al., 2023). This comprehensive dataset enables me to identify employees in federal states with and without collective bargaining compliance laws.

The SIAB has some limitations, which will be addressed as follows. First, the SIAB does not indicate whether a company has a public service contract or not. To address this, I restrict the data to employees and establishments in the public transport sector. The public transport sector is inherently organized by state or local authorities and operated by both public and private companies. This means all establishments function on behalf of the state. Public transport services are legally defined as those provided by trams, buses, and motor vehicles in regular service, primarily for urban, suburban, or regional transport. Thus, I can classify local public transport establishments using the five-digit economic classification level since 2008, retroactively applying this information to earlier years for these establishments.[10]

Second, the SIAB does not contain information on working hours. Therefore, I restrict my sample to full-time employees aged 18 to 65, consistent with other wage analyses using the SIAB (e.g. Baumgarten et al., 2020; Riphahn & Schnitzlein, 2016). Employees with multiple jobs are excluded, as their working hours may adjust following the introduction of the law. Additionally, temporary workers are omitted because they are paid by temporary employment agencies, rendering the impact of the collective bargaining compliance law on their wages unclear. Furthermore, to approximate the conditions of an ideal experiment and mitigate self-selection effects, I exclude employees who have changed their federal state of employment. This approach ensures that the analysis is concentrated on employees who have remained in the same federal state throughout the study period, thereby minimizing potential selection bias.

---

[10] The economic sector is defined using Local passenger transport on land (excluding taxis) (49310) from the economic classification w08. The imputation procedure for a time-constant recording of economic sectors at the five-digit level, developed by Drechsler and Ludsteck (2024), is not yet available from the IAB Research Data Centre at the time of my analysis.

Third, the SIAB does not provide information whether a company is bound by collective agreements.[11] However, this limitation is of minor concern, as the law's application is not dependent upon collective bargaining coverage at the establishment level. Establishments with a collective agreement that pay less than the level mandated by the representative collective agreement required by the federal state are still subject to the law.

For my outcome variable of interest, the imputed daily wage, I impose the following restriction: I exclude full-time employees with a daily wage below EUR 30. I consider such low wages for full-time employees to be unrealistic and likely to indicate data inaccuracies. While this threshold is admittedly somewhat arbitrary, it affects less than one per cent of the observations. Moreover, in some federal states, a sector collective agreement exists for employees in the public transport sector, where the lowest wage group starts with a daily wage of more than EUR 40.[12] In view of this fact, the limit of EUR 30 per day for full-time employees is justifiable.

With regard to the introduction of laws in the individual federal states, I apply the following data restriction to enable a comparison between states with and without collective bargaining compliance laws. I exclude federal states that introduced and then repealed collective bargaining compliance laws before and up to 2008.[13] This is crucial because public service contracts are time-limited, with maximum durations of ten years for bus services and 15 years for rail services (European Union, 2007). Additionally, laws became more consistent across states following the European Court of Justice rulings in 2008. In 2008, the year of the Rüffert ruling, it was uncertain whether public transport laws were applied contractually, so I focus on post-2008 laws where the legal standing is clearer.

---

[11] The IAB provides the Linked Employer-Employee Data (LIAB), a dataset that links information from the SIAB with the IAB Establishment Panel. The IAB Establishment Panel is an annual representative employer survey on the determinants of employment and includes information about a company's collective bargaining status. However, the IAB Establishment Panel is not representative at the five-digit economic classification level used in my analysis. Consequently, the LIAB cannot be utilised due to a lack of representativeness (Panahian Fard et al., 2024).

[12] Federal states: Bavaria, Berlin, North Rhine-Westphalia, Baden-Württemberg and Lower Saxony apply the local public transport collective agreement for public transport employees: https://oeffentlicher-dienst.info/tv-n/.

[13] Berlin, Bremen, Hamburg, Lower Saxony, North Rhine-Westphalia and Schleswig-Holstein are excluded.

The control group is handled as follows during data preparation. The dataset includes two federal states, Bavaria and Saxony, which never implemented a collective bargaining compliance law and thus serve as control groups. Schnabel (2016) recommends analysing West and East Germany separately due to differences in bargaining coverage and other labour market characteristics. Furthermore, Oberfichtner and Schnabel (2019) demonstrate that both parts of Germany differ substantially in terms of collective bargaining coverage and worker representation through works councils, which are the two main pillars of the German industrial relations model. Therefore, I analyse West and East Germany separately. Bavaria serves as the control for Western states, while Saxony serves as the control for Eastern states. Considering ongoing discussions on choosing control groups for analysing staggered interventions (Callaway & Sant'Anna, 2021; Roth et al., 2023), I create two datasets to ensure a precise selection of control observations.

The first dataset is used for descriptive analysis and primary results based on a standard dynamic two-way fixed effects (TWFE) event study. In this dataset, I replicate the data from Bavaria and Saxony for each law's implementation[14] in other states, assigning each replication an artificial event year corresponding to its control group's event year. Observations are then balanced according to these assigned event years, allowing for continuous tracking of individuals in both the treatment and control groups from seven years before to five years after the (artificial) event. This ensures clear comparisons for each year of the law's enactment. Consequently, my estimated effect is a pooled result from individual event studies. Figure 1 depicts the distribution of employees subject to social security contributions in the public transport sector in West and East Germany, categorized by the law implementation year and treatment status. It reveals more comparable numbers of observations between the control and treatment groups in East Germany. Given that Saarland (implementation in 2010) and Rhineland-Palatinate (implementation in 2011) in West Germany are smaller federal states compared to Bavaria or Baden-Württemberg, this result is unsurprising.

(Figure 1 around here)

---

[14] For stylistic reasons, this study uses "implementation" and "enactment" interchangeably to refer to the year the laws are executed

Event studies using standard TWFE estimators have been significantly critiqued in recent literature (Roth et al., 2023). This criticism stems from these estimators providing reliable estimates only if treatment effects are homogeneous, regardless of when laws are implemented. If effects vary over time among states, TWFE estimators can produce biased coefficients due to negative weights. Therefore, I prepare a second dataset to apply the Callaway and Sant'Anna (2021) estimation, which avoids the issue of negative weights and offers more robust results (Callaway & Sant'Anna, 2021; Roth et al., 2023).

This second dataset follows the same data restrictions as the first, but it does not replicate data from Bavaria and Saxony, avoids assigning artificial event years, and balances observations on an annual basis. In this dataset, individuals are balanced across calendar years instead of relative to the year of law enactment, allowing employees to be consistently observed from 2000 to 2020, rather than only seven years before to five years after an event.

## 5. Descriptive evidence

In Table 3, average imputed daily wages are presented separately for West and East Germany, divided by treatment group and period. In West Germany, during the period from seven years to one year before the law's introduction, wages in the treated federal states average EUR 108.69 per day, which is EUR 3.34 higher than the control group, Bavaria, where wages average EUR 105.35. This wage difference is statistically significant at the one per cent level. Following the law's introduction, wages in the treatment group increase by 10.2 per cent, reaching an average of EUR 119.78. Over the same period, wages in Bavaria rise by 7.4 per cent to EUR 113.17 per day. This results in a daily wage difference of EUR 6.61, which is statistically significant at the one per cent level.

In East Germany, the ratio of average wages before the implementation of the collective bargaining compliance law is reversed compared to West Germany. Table 3 shows that during the period from seven years to one year before the law's enactment, daily wages in the treated federal states average EUR 90.54, which is EUR 3.25 lower than the control group, Saxony, at EUR 93.79. This difference is statistically significant at the one per cent level. After the law's introduction, wages in the treated federal states rise by 7.7 per cent to reach EUR 97.48, while Saxony sees a 4.6 per cent increase to EUR 98.14. Notably, the average wage difference between the groups narrows to just EUR 0.66, which is neither economically nor statistically significant.

(Table 3 around here)

While Table 3 offers an initial glimpse of the potential impacts of collective bargaining compliance laws, it does not specify when the observed wage increases begin. Figure 2 depicts the development of wages over the specified period - from seven years before to five years after the law's implementation - separately for West and East Germany, and by treatment status.

In West Germany, wages for public transport employees in treated federal states with a compliance law are higher than in the control group, Bavaria. However, a closer look reveals that the wage increase in treated states begins two years prior to the law's enactment, at t-2, rather than at the time of its introduction. Until t-2, wages in the treated western states and Bavaria develop nearly parallel. The rise starting at t-2 suggests divergence in wage development between these treated states and Bavaria, raising questions about the validity of the parallel trend and no anticipation assumptions in the absence of the law, which is crucial for properly assessing the law's effects in subsequent econometric analysis.

In East Germany, throughout the entire period, wages in the control group, Saxony, are higher than in the treated states. However, before the law's introduction, wage developments in both the treated and control groups follow nearly parallel trends. After the law's enactment, the treated East German states experience a more pronounced wage increase, bringing their wages on average to the level of those in Saxony. Consequently, in the year following the law's introduction, wages between the treated and control groups nearly converge, aligning with the average wage results observed in the pre- and post-law periods, as shown in Table 3.

(Figure 2 around here)

In summary, the descriptive analyses indicate that collective bargaining compliance laws are associated with wage increases, in East Germany, whereas in West Germany, wage increases were apparent even before the law was enacted. This suggests a potential violation of the parallel trend and no anticipation assumptions in West Germany, while these assumptions remain intact in East Germany.

## 6. Empirical model

The objective of this paper is to explore the relationship between collective bargaining compliance laws and wage development for employees in the public transport sector. The introduction of the laws across federal states can be considered quasi-random and thus unlikely to be correlated with firm- or individual-level characteristics. In my first specification, I use a standard dynamic two-way fixed effects (TWFE) event study estimator as described by Roth et al. (2023). The basic idea is to compare the wages of individuals with similar characteristics who differ only by their federal state. Assuming no anticipation effects and parallel trends - meaning wages weren't adjusted prior to the laws and would have evolved similarly in states with and without the collective bargaining compliance laws - any observed wage changes can be attributed to the law's implementation. To illustrate changes over time, I use an event study model structured as follows:

$$y_{i,t} = \sum_{\tau \neq -1} \beta_\tau I(t = \tau) + \sum_{\tau \neq -1} \gamma_\tau I(t = \tau) I(T_i = 1) + \alpha_i + \delta_c + \phi_e + \epsilon_{i,t}$$

where $y_{i,t}$ is the log of the deflated imputed daily wage of individual $i$ at time $t$, and $T_i$ is a treatment dummy that equals 1 if the individual works in a federal state with a collective bargaining compliance law and 0 otherwise. $\alpha_i$ captures individual fixed effects, $\delta_c$ represents calendar year fixed effects using year dummies, and $\phi_e$ accounts for establishment fixed effects by including establishment dummies. $\epsilon_{i,t}$ represents an idiosyncratic error term.

For a fixed time, point $\tau$, $\beta_\tau$ captures the average value of outcomes for the control group relative to the reference period (conditional on fixed effects), while $\gamma_\tau$ represents the average difference between the treatment and control groups at that specific time point. The variable $\tau$ represents the event time and ranges from $t-7$ to $t+5$, covering a 13-year period. The event-time values less than or equal to $t-6$ are combined into a single dummy to avoid collinearity, as described in Miller (2023). Following standard practice, I use the event time dummy $t-1$ as a reference group. Abadie et al. (2022) emphasize that the selection of the cluster unit for the standard errors should be aligned to the experimental design of the respective study. Standard errors are therefore clustered at federal state level since the laws are introduced at the federal state level.

In my analysis of both West and East Germany, I include only five federal states in each region. This small number of clusters violates a key assumption for calculating clustered standard errors, which is that the number of clusters should be large. When this assumption is violated, it can lead to an over-rejection of the null hypothesis. To address this issue, I calculate p-values based on wild bootstrap errors for the TWFE estimator results. Roth et al. (2023) critically note that the wild bootstrap can perform well in certain scenarios with a small number of clusters but it too requires strong homogeneity assumptions, especially when a dynamic TWFE estimator is applied. Ultimately, the issue of having too few clusters cannot be fully resolved, but applying different approaches to verify the validity of my inference should help assess the robustness of my conclusions.

Event studies using standard dynamic TWFE estimators have been widely criticized in recent literature (e.g., Borusyak et al., 2024; de Chaisemartin & D'Haultfœuille, 2020; Goodman-Bacon, 2021; Roth et al., 2023; Sun & Abraham, 2021) due to their limitations when evaluating staggered interventions, such as the enactment of collective bargaining compliance laws at different times across various federal states. Dynamic TWFE estimators provide reliable estimates only if the treatment effect is homogeneous, regardless of when these laws were enacted by the states. However, when the effects are heterogeneous over time across states, dynamic TWFE might result in biased coefficients (Roth et al., 2023). This is because the TWFE method calculates law enactment effects as a weighted sum of average treatment effects (ATT), where negative weights can produce misleading results. These negative weights can arise due to the so-called "forbidden comparisons," where early-treated units are used as a control group for later-treated units. Consequently, dynamic TWFE can present negative coefficients for the ATTs even if the ATTs are positive (Roth et al., 2023). Roth et al. (2023) highlight the implication from Sun and Abraham (2021) that if treatment effects are heterogeneous, the coefficients from the period before law implementation may not be zero, even if parallel trends exist across all periods. Thus, evaluating pre-trends based on these coefficients can be misleading (Roth et al., 2023). This must be considered when using dynamic TWFE estimations to evaluate policies.

To address the issues with the TWFE estimator, various studies have proposed alternative estimators that can sensitively detect aggregate heterogeneous treatment effects in settings with staggered treatment timing (e.g. Callaway & Sant'Anna, 2021; de Chaisemartin & D'Haultfœuille, 2020; Sun & Abraham, 2021; Wooldridge, 2021). These estimators vary primarily in their identifying assumptions, comparison groups, and efficiency characteristics (see, for example, the overview articles by de Chaisemartin and D'Haultfœuille (2023) Chaisemartin and D'Haultfoeuille (2023) and Roth et al. (2023)).

Due to the limitations of the standard TWFE method, I employ the estimator proposed by Callaway and Sant'Anna (2021) as a second specification in my analysis. This estimator has been applied in recent studies with staggered interventions (e.g., Beeson et al., 2024; Peña & Sanso-Navarro, 2025; Pham et al., 2025; Robertson, 2023).

In contrast to dynamic TWFE, this estimator identifies the average treatment effect on the treated (ATT) at a specific time ($t$) relative to the time of treatment (g), under staggered versions of the parallel trends and no anticipation assumptions. It does so by comparing the expected change in outcomes for the treated cohort ($g$) between periods ($g-1$) and ($t$) to the change for an appropriate control group, such as units not yet treated by period ($t$), as intuitively explained by Roth et al. (2023).

In other words, the Callaway and Sant'Anna (2021) estimator computes a series of ATTs for different years relative to the year of introduction, assuming parallel trends between treatment and control groups only between two specific time points, rather than over a broader period. For example, the Callaway and Sant'Anna (2021) estimator would calculate an ATT(2015,2017)[15], meaning the average treatment effect in 2017 for federal states that enacted the collective bargaining compliance law in 2015, under the assumption that the parallel trend assumption holds only between these points in time. This calculation is repeated for various ATTs, such as ATT(2015,2018), ATT(2015,2019), and so on. These individual ATTs can then be summarised as a (weighted) average of the treatment effect to obtain event-study coefficients (see Callaway and Sant'Anna (2021)).

According to Roth et al. (2023), Callaway and Sant'Anna (2021) estimation offers two advantages over dynamic TWFE estimation. First, it provides sensible estimates even with arbitrary heterogeneity of treatment effects by avoiding negative weighting, which is not necessarily the case with TWFE. Second, the estimator by Callaway and Sant'Anna (2021) clearly shows which units are used as a control group to infer the unobserved potential outcomes.

---

[15] This example is based on the explanation in Roth et al. (2023).

I implement the Callaway and Sant'Anna (2021) estimator using two approaches. Initially, I apply its standard form without establishment fixed effects, focusing on person and year fixed effects. My main specification, however, includes establishment fixed effects using establishment dummies to match the dynamic TWFE approach. I calculate uniform confidence intervals with a multiplicative wild bootstrap, as described by Callaway and Sant'Anna (2021). These simultaneous confidence bands, unlike traditional pointwise intervals, cover the full path of group-time average treatment effects with fixed probability and account for dependencies between estimators. According to Callaway and Sant'Anna (2021), and as noted by Roth et al. (2023), these confidence bands provide a potentially more robust inference method and are arguably more appropriate than traditional pointwise confidence intervals.

## 7. Results and discussion
### 7.1 Main results

Figure 3 displays the results of the dynamic Two-Way Fixed Effects (TWFE) event study model alongside the findings from the Callaway and Sant'Anna (2021) estimator, which are presented both with and without establishment fixed effects. The coefficients represent the wage differential between treated and never-treated states, both before and after the enactment of the bargaining compliance law, relative to the reference period (t-1) for both West and East Germany. As highlighted by Miller (2023), the coefficients from the pre-law period are instrumental in assessing the validity of the parallel trends assumption.

(Figure 3 around here)

In West Germany, the dynamic TWFE event study reveals consistently negative pre-period coefficients that are significantly different from zero, ranging between 4.8 and 3.8 per cent within the five years prior to the law's enactment. However, according to Sun and Abraham (2021) and Roth et al. (2023), caution is advised when evaluating the parallel trends assumption based on TWFE, as the TWFE estimates can be misleading due to potential negative weighting in the pre-trends. Looking at the Callaway and Sant'Anna (2021) estimation, which does not account for establishment fixed effects and avoids the issue of negative weighting, reveals similar negative point estimates to the TWFE estimation, but most coefficients are not statistically significantly different from zero. Only the point estimate for the period t-2 is statistically significant, indicating a negative difference of 4.2 per cent. This may suggest anticipation and thus a violation of the no anticipation and parallel trend assumptions, supporting the descriptive insight from Figure 2. When accounting for establishment fixed effects in the Callaway and Sant'Anna (2021) estimation, no coefficient in the pre-period is statistically significantly different from zero. The point estimates range from plus 0.2 to minus 1.6 per cent within five years before enactment, indicating they are also economically insignificant. Thus, the validity of the parallel trend and no anticipation assumptions can be assumed.

Interestingly, all three estimates for West Germany in the years following the enactment of the law show very similar point estimates, ranging between minus 2.9 and plus 0.7 per cent. These coefficients are statistically insignificant and can also be considered economically insignificant. The results from the TWFE and Callaway and Sant'Anna (2021) estimations without establishment fixed effects should be interpreted with caution, as the central assumptions of no anticipation and parallel trends are violated. This raises questions about whether Bavaria is an appropriate control group for the other western states. However, when establishment fixed effects are included in the Callaway and Sant'Anna (2021) estimation, the assumptions hold, suggesting that the collective bargaining compliance law had no effect in the western federal states. Although it is plausible that the existence of sectoral agreements in public transport in the western states prevented the collective bargaining compliance laws from creating an additional wage impact, this cannot be empirically tested here. For future evaluations of collective bargaining compliance laws, it would be interesting to examine how the presence of sectoral agreements influences the effectiveness of these laws.

In East Germany, both the TWFE estimation and the Callaway and Sant'Anna (2021) estimation with establishment fixed effects show no statistically significant differences in wage development between the treated federal states and the control group, Saxony, during the entire pre-law period. Only, the Callaway and Sant'Anna (2021) estimation without establishment fixed effects reveals one statistically significant coefficient, showing a negative 3.7 per cent difference four years before enactment, while all other point estimates remain statistically insignificant. Despite this one coefficient, the overall estimations from both TWFE and Callaway and Sant'Anna (2021) estimations suggest that the assumptions of parallel trends and no anticipation is satisfied. This implies that, in the absence of collective bargaining compliance laws, wage development in the treated states would likely have mirrored that of Saxony. Additionally, most coefficients in all three estimations are negative, confirming the findings from Figure 2 that, prior to the enactment of the laws, wages in the treated federal states were lower than in Saxony.

In the year the law was enacted, wages increased by an average of 2.0 per cent according to the TWFE estimation, 2.2 per cent according to the Callaway and Sant'Anna (2021) estimation without establishment fixed effects, and 2.7 per cent according to the Callaway and Sant'Anna (2021) estimation with establishment fixed effects. This immediate effect is likely due to the time gap between the law's passage and enactment. All three estimations show a similar pattern for up to five years post-enactment. The TWFE estimation indicates that wage increases remained consistently high during this period, with a 3.2 per cent rise one year after enactment, persisting at 3.4 per cent by the fifth year, with all coefficients statistically significant at the five per cent level. Similarly, the Callaway and Sant'Anna (2021) estimation without establishment fixed effects identifies a significant 3.3 per cent wage difference in the first year post-enactment, remaining steady at 2.9 per cent by the fifth year, though statistically insignificant. Using the same fixed effects as in the TWFE, the Callaway and Sant'Anna (2021) estimation including establishment fixed effects provides slightly higher results, demonstrating a 3.7 per cent wage difference in the first year and a 4.6 per cent difference by the fifth year following enactment. All coefficients of this estimation remain significant at the five per cent level, similar to the TWFE estimation. Overall, the coefficients between the different estimations do not differ statistically, as seen from the overlapping confidence intervals.

In summary, the results for East Germany are not only statistically significant but also economically meaningful. In contrast, in West Germany, when including establishment fixed effects in the Callaway and Sant'Anna (2021) estimation, no effects of the law's introduction are found. The differences between West and East Germany could possibly be attributed to the generally higher collective bargaining coverage in West Germany and the presence of sectoral collective agreements in public transport, such as in Bavaria and Baden-Württemberg. These sectoral agreements are absent in East

Germany, which, on one hand, increases the comparability between control and treatment groups and, on the other hand, could explain why the TWFE estimation and the Callaway and Sant'Anna (2021) estimation without establishment fixed effects in West Germany cast doubt on the comparability between Bavaria and the other western federal states.

The observed wage effects, particularly for East Germany can be interpreted as an "intention-to-treat" effect. Thus, the observed effect should be understood as a lower bound and might be larger if all individuals in public transport were executing contracts awarded under the collective bargaining compliance law. This is because the data does not allow me to observe whether or when a public contract subject to a collective bargaining compliance law was awarded. These limitations lead to a situation where both treated and untreated individuals should coexist in states with such laws. This makes interpreting the effect as an average treatment effect not possible. If the analysis captured only treated individuals, the estimated effect might be stronger. The magnitude of my effect, up to at least 4.6 per cent, aligns with existing literature on collective bargaining premiums, which show a wage premium of 1 to 4 per cent (Addison et al., 2014; Bonaccolto-Töpfer & Schnabel, 2023; Hirsch & Mueller, 2020; Jäger et al., 2024). Similar wage increases were observed following the introduction of the minimum wage in Germany, ranging from 4.8 to 6.5 per cent (Bossler & Gerner, 2019; Burauel et al., 2020). Moreover, my results are comparable with the wage effects of living wage introductions, which increase wages within establishments by an average of 4.1 to 4.3 per cent in the UK (Datta & Machin, 2024).

The central question remains whether the observed effect in East Germany can be interpreted causally. Although the parallel trend and no anticipation assumptions are satisfied, the data do not provide crucial details on whether and when establishments receive public contracts, nor do they include information on contract size, the extent of collective bargaining coverage, or municipalities' decisions to bypass public procurement laws through direct contract awards. The causal inference hinges on the assumption that unobserved, time-varying factors influencing the treatment effect and wages do not systematically differ between states with and without compliance laws. If valid, differences between treatment and control groups would account for these unobserved variables, making my results representative of a causal effect. Given the uniform legal framework and population-based provision of public transport, which is fairly consistent across all eastern federal states (Figure 4), significant systematic differences seem improbable. However, this cannot be decisively established; if such differences exist, the estimated effects would likely reflect correlation rather than causation.

(Figure 4 around here)

Despite the limitations and unresolved questions of causality, the findings suggest that collective bargaining compliance laws are at least correlated with increases in employee wages. As federal states in Germany are major public spenders (Statistisches Bundesamt (Destatis), 2025), the analysis of state-level legislation provides initial insights into how a federal law might be associated with wage rises. This insight is particularly relevant in political discussions, highlighting how state laws can enforce collectively agreed wages in the labour market in a setting — such as East Germany — where collective bargaining coverage is lower, and no sector-wide agreements exist for the sector in question. However, these findings must be contextualised: the public transport sector is dominated by state demand, making it challenging for companies to replace state contracts with private alternatives, unlike other sectors. Additionally, the results for West Germany suggest that in areas with existing sectoral agreements, collective bargaining compliance laws have little impact. It remains uncertain whether similar wage effects would be observed in sectors where public contracts can be more easily substituted with private ones or in industries where sectoral agreements or industry-specific minimum wage regulations are more prevalent. Therefore, before the implementation of a collective compliance law, the federal legislator should consider whether the federal government is awarding contracts in industries where collective bargaining coverage is weak and whether there are no other wage-setting mechanisms, such as industry-specific minimum wages, in place.

## 7.2 Discussion of inference

The standard errors of the TWFE estimation are clustered at the federal state level, which is the level of law implementation. This has the drawback that only five clusters are available in the estimate for both West and East Germany. This low number of clusters leads to a violation of the key assumption for calculating clustered standard errors, namely that the number of clusters is large. Having a low number of clusters leads to an over-rejection of the null hypothesis (Cameron et al., 2008; Djogbenou et al., 2019; MacKinnon & Webb, 2017, 2018). This section presents alternative approaches to clustering in order to test the robustness of the main results from my dynamic TWFE event study.

Cameron et al. (2008) point out that over-rejection of the null hypothesis can occur when only five to ten clusters are available. In such cases, the authors recommend using the wild cluster bootstrap method to eliminate bias. However, MacKinnon and Webb (2017) document that in settings with fewer treated (or untreated) clusters, the desired properties of the wild cluster bootstrap are not maintained: the wild cluster bootstrap with restricted heteroscedasticity (WCR) under-rejects, while the wild cluster bootstrap with unrestricted heteroscedasticity (WCU) over-rejects the null hypothesis. In such cases, MacKinnon and Webb (2018) suggest clustering at a finer level to improve the reliability of inference. To test if the desired properties of the wild cluster bootstrap are violated, Roodman et al. (2019) suggest estimating both the WCR and WCU and comparing the results. Wild cluster bootstrapping may be problematic if inferences from the two methods differ.

To check the validity of my results, I follow Roodman et al. (2019) and calculate both versions of the wild cluster bootstrap. First, I cluster the bootstrap errors by state, and second, I cluster them by individual - the finest level in my data. As described by MacKinnon and Webb (2018) and mentioned by Roodman et al. (2019), the latter approach represents the ordinary wild bootstrap procedure. Table 4 present the results for West and East Germany, respectively. It shows the estimated coefficients from Figure 3 with standard errors clustered at the federal state level in parentheses, as well as the calculated p-values based on wild bootstrap errors clustered by state and by individual.

(Table 4 around here)

The results for West Germany (Table 4) show that the WCR method reports a higher significance level than the WCU method. This difference arises because WCR under-rejects the null hypothesis, while WCU over-rejects it. When clustered by individual, WCR and WCU results nearly coincide. Despite differences at the WCR and WCU clustered at the federal state level, the fundamental conclusions do not contradict each other. Across all calculations, the coefficients for the pre-treatment period are statistically significantly different from zero, despite differences in significance levels. This reinforces the finding that the parallel trends assumption is violated in West Germany. Therefore, a valid statement about wage development between Bavaria and the western treated federal states cannot be made using TWFE estimation.

The results for East Germany (Table 4) indicate that the coefficients during the pre-treatment period are not statistically significantly different from zero, regardless of the wild cluster bootstrap method used. This finding supports the conclusion that the parallel trends assumption is valid in this context. In the post-policy period following the law's introduction, there is a slight difference in significance levels between the WCR and WCU methods when clustered by state: WCR rejects the null hypothesis at the 5 per cent level, whereas WCU rejects it at the 1 per cent level. This pattern is not observed for the coefficient at event time +4, where WCU rejects the null hypothesis at a higher significance level than WCR. The only conflicting result is for the coefficient at event time +2. Here, WCR does not reject the null hypothesis, whereas WCU rejects it at the 5 per cent level. When clustered by individual, the results from both wild cluster bootstrap methods are nearly identical. Comparing the results of clustering by state versus individual, the findings consistently show that the null hypothesis can be rejected at the 5 per cent level. The sole exception is the coefficient at event time +2, which is not significant when clustered by WCU at the state level, but is significant with all other methods.

Even though Roth et al. (2023) critically observe that the wild bootstrap can be effective in certain scenarios with a small number of clusters, it necessitates strong homogeneity assumptions, particularly when a dynamic TWFE estimator is used. The results demonstrate that, despite this minor difference, the conclusion remains valid regardless of whether wild bootstrap or clustering standard errors at the federal state level are used: In East Germany, the parallel trend assumption holds, and wages significantly increased, whereas in West Germany, the parallel trend assumption does not hold, casting doubt on the validity of the results for West Germany when using TWFE.

### 7.3. Robustness checks

The presented results may be influenced by three confounding factors: firstly, the introduction of the minimum wage in Germany in 2015; secondly, the influence of certain federal states that may disproportionately drive the overall effect; and thirdly, other unobserved factors. To address these issues, I conduct three different robustness checks for the standard case of the dynamic TWFE estimation and my preferred model, the Callaway and Sant'Anna (2021) estimation, including establishment fixed effects:

Firstly, the introduction of the minimum wage in Germany in 2015, set at EUR 8.50, falls within the observation period and may influence the observed effects, potentially attributing them to the minimum wage rather than the collective bargaining compliance law. Data from the IAB Establishment Panel, as noted by Bossler and Gerner (2019), show that businesses in East Germany were notably impacted, with a higher proportion of employees earning below EUR 8.50 before the implementation of the generally binding minimum wage compared to West Germany. To evaluate the robustness of the findings, the main dynamic TWFE event study analysis is re-run, excluding daily wages below EUR 68.00, which corresponds to the daily wage of a full-time employee working an eight-hour shift at the minimum wage rate. Comparing the TWFE estimates and the Callaway and Sant'Anna (2021) estimations with establishment fixed effects from the main analysis to those after excluding these lower wages (Figure A1), shows that the results for both West and East Germany remain nearly identical in size and significance. This indicates that the findings are not driven by the introduction of the minimum wage.

Secondly, I investigate whether any individual federal states might influence my results. To achieve this, I repeat the TWFE estimation and Callaway and Sant'Anna (2021) estimations with establishment fixed effects, sequentially excluding one treated federal state at a time. Comparing the main results with those from this robustness check shows that the estimates remain consistent. The coefficients for both West (Figure A2) and East Germany (Figure A3) are not significantly different from those in the main model, and the relationships identified in the main analysis hold overall. Only in the Callaway and Sant'Anna (2021) estimation for East Germany does the exclusion of Mecklenburg-Western Pomerania suggest a potential violation of the parallel trend assumption since the coefficients in the periods four and three years before enactment are statistically significant. However, the coefficients lie within the confidence interval of the main specification, so they are not statistically significantly different from each other. Therefore, my main results are not caused by the observations from Mecklenburg-Western Pomerania. Moreover, the point estimates for the period after the law's enactment are not significantly different from those of the main specification, even when Mecklenburg-Western Pomerania is excluded. Therefore, I would argue that my main results are robust against influential observations.

Thirdly, as an additional robustness check, a placebo policy implementation year is introduced by moving the actual introduction year forward by three years. Federal states that implemented the law in 2010 are reassigned to 2007, those with a 2011 introduction year to 2008, those with a 2012 introduction year to 2009, and so on. Apart from this change, the data were prepared exactly as in the main datasets, with balancing over the event year or calendar years. This approach tests whether the actual introduction year is the main factor driving the observed effects. The results show that no effects consistent with the main findings are detected across both regions using the placebo policy years, thereby confirming that the actual introduction year is the primary factor driving the observed effect and confirming the robustness of the main results.

In West Germany, neither the TWFE estimation nor the Callaway and Sant'Anna (2021) estimation shows any effect: almost all coefficients are statistically insignificant, and their size, particularly in the Callaway and Sant'Anna (2021) estimation, is close to zero. In East Germany, a significant positive effect is evident only from period t+3 onwards in the Callaway and Sant'Anna (2021) estimation, and a similar increase is observed in the TWFE estimation, although the coefficients are not significant in this case. The increase from period t+3 aligns with the original policy introduction year. All other coefficients do not differ significantly from each other, confirming the robustness of the findings in East Germany (Figure A4).

In conclusion, the findings show that the Collective Bargaining Compliance Law increased wages for public transport employees in East Germany. However, in West Germany, the Callaway and Sant'Anna (2021) estimation with establishment fixed effects shows no significant impact. Conversely, the TWFE estimation and the Callaway and Sant'Anna (2021) estimation without these fixed effects raise doubts about the comparability between the treatment and control groups, preventing a clear assessment of wage effects.

## 8. Conclusion

This study provides novel insights into the wage effects of collective bargaining compliance laws (Tariftreuegesetze) in Germany. These laws mandate that companies must pay wages and ensure working conditions according to a representative collective agreement, even if they are not directly bound by such an agreement with a trade union. Using administrative data and examining the implementation of these laws at the federal state level, event study estimates reveal that wages for public transport employees increased by 2.9 to 4.6 per cent within five years of the law's implementation—an effect observed exclusively in East Germany. In contrast, no wage effects were detected in West Germany based on Callaway and Sant'Anna (2021) estimations, including establishment fixed effects. These differences between West and East Germany could be possibly attributed to the generally higher collective bargaining coverage and the presence of sectoral collective agreements for public transport in West Germany, although this cannot be empirically tested within the scope of this analysis. Nevertheless, from an econometric perspective, East Germany offers a more suitable setting for the analysis, as the eastern federal states are institutionally homogeneous due to the absence of sectoral collective agreements. This institutional consistency comes closer to the conditions of a natural experiment.

Regarding the question of causality, the estimated effect should not be interpreted as an average treatment effect but rather as an "intention-to-treat" effect. In the data I cannot observe which individuals are treated, leading to a situation where both treated and untreated individuals should coexist in states with collective bargaining compliance laws. Therefore, econometrically, the observed wage effects represent a lower bound and should be larger if all individuals were treated. Notably, all results remain robust even when considering the implementation of the generally binding minimum wage in Germany in 2015.

My findings contribute to and align with existing literature on collective bargaining premiums (Addison et al., 2014; Bonaccolto-Töpfer & Schnabel, 2023; Hirsch & Mueller, 2020; Jäger et al., 2024), as well as the documented effects of minimum wages in Germany (Bossler & Gerner, 2019; Burauel et al., 2020). They also correspond with studies on living wages in the UK (Datta & Machin, 2024). Additionally, this study is the first econometric analysis to engage with the ongoing political and legal discourse regarding the adoption of a collective bargaining compliance law at the federal level in Germany (Caspers, 2024; Schulten, 2021).

The study is constrained by data limitations when analysing the wage effects. The dataset lacks crucial information on whether a company applies for or holds a public contract, and it is missing information on the timing of contract awards. To mitigate these issues, the study focuses specifically on the public transport sector. In Germany, public transport is state-organised, meaning that almost all companies within this sector hold a state contract. Although I cannot fully resolve the issue of determining the timing of these awards, I argue that within five years of the law's enactment, some municipalities in Germany should grant new contracts for their public transport services. Moreover, municipalities have the option to award contracts directly, which are not subject to the collective bargaining compliance law; however, this option is restricted by EU law. This situation is further addressed econometrically through two-way fixed effects (TWFE) and the Callaway and Sant'Anna (2021) event studies. Both methods examine differences between treatment and control groups. Given the lack of evidence indicating significant differences in awarding practices between municipalities in treated and control states, these differences help address the issue of direct awards.

Despite its limitations, this study provides valuable insights into how collectively agreed wages can be secured in the context of declining collective bargaining coverage and offers an early indication of the potential effects of a federal collective bargaining compliance law. However, the implications must be carefully contextualised to guide political action. The analysis is centred on a sector where the state possesses substantial market power, with state contracts not easily substitutable by private ones. It remains uncertain whether similar wage effects would emerge in sectors where public contracts could be more easily replaced by private alternatives or in sectors where sector-specific collective agreements or sector-specific minimum wages are in play.

Policymakers should consider adjustment mechanisms in establishments when creating legislation to ensure the state's market power can make these laws effective. It is important to assess whether and to what extent the federal government awards contracts in areas with low collective bargaining coverage or no sector-specific minimum wages. This is crucial because collective bargaining compliance laws are most effective in such settings. From both economic and legal viewpoints, enhancing the process of making collective agreements generally binding may be more effective. This approach, embedded in German labour law, reduces adjustment mechanisms and ensures that the benefits of collectively agreed wages extend beyond those employees fulfilling state contracts since general binding collective agreements affect both private and public contracts.

Table 1: Adoption, Enforcement, Abolishment, and Thresholds of Collective Bargaining Compliance Laws in Public Transport Across Federal States in Germany

| Federal States | Law Adoption Date: | Law Enactment Date: | Law Abolition Date: | Threshold Value in Euros by introduction |
|---|---|---|---|---|
| **West Germany** | | | | |
| Till 2008 | | | | |
| Bremen | 17 December 2002 | 01 March 2003 | | 10,000 |
| Lower Saxony | 02 September 2002 | 01 January 2003 | 31 December 2005 | 10,000 |
| | 31 October 2013 | 01 January 2014 | | 10,000 |
| Schleswig-Holstein | 07 March 2003 | 28 March 2003 | 31 October 2010 | 10,000 |
| | 31 May 2013 | 01 August 2013 | | 15,000 |
| North Rhine-Westphalia | 17 December 2002 | 01 March 2003 | 20 November 2006 | 10,000 |
| | 10 January 2012 | 01 May 2012 | | 20,000 |
| Hamburg | 18 February 2004 | 01 April 2004 | 31 December 2008 | - |
| After 2008 | | | | |
| Saarland | 15 September 2010 | 05 November 2010 | | 50,000 |
| Rhineland-Palatinate | 01 December 2010 | 01 March 2011 | | 20,000 |
| Baden-Württemberg | 16 April 2013 | 01 July 2013 | | 20,000 |
| Hesse | 19 December 2014 | 01 March 2015 | | 10,000 |
| Bavaria | No adoption of a collective bargaining compliance law | | | |
| **East Germany** | | | | |
| Till 2008 | | | | |
| Berlin* | 09 July 1999 | 30 March 2008 | | - |
| After 2008 | | | | |
| Mecklenburg-Western Pomerania | 07 July 2011 | 16 July 2011 | | - |
| Thuringia | 18 April 2011 | 01 May 2011 | | 20,000 |
| Brandenburg | 21 September 2011 | 01 January 2012 | | 3,000 |
| Saxony-Anhalt | 19 November 2012 | 01.January 2013 | | 25,000 |
| Saxony | No adoption of a collective bargaining compliance law | | | |

Note: The table shows the adoption, enactment, abolishment, and threshold values of collective bargaining compliance requirements concerning the public transport sector across federal states in West and East Germany, both before and after the European Court ruling in the Rüffert case in 2008.*The Berlin Procurement Law of 9 July 1999 applied initially to construction services and services related to buildings and real estate, and was expanded on 13 March 2008 to encompass all sectors, including public transport. This sectoral extension took effect on 30 March 2008.

| Table 2: Percentage of Tenders Referencing Direct Award in Title and/or Description ||||
|---|---|---|---|
| Year | No | Yes | N |
| 2011 | 100.00 | 0.00 | 29 |
| 2012 | 97.50 | 2.50 | 40 |
| 2013 | 97.80 | 2.20 | 91 |
| 2014 | 98.20 | 1.80 | 111 |
| 2015 | 95.50 | 4.50 | 111 |
| 2016 | 99.14 | 0.86 | 116 |
| 2017 | 99.06 | 0.94 | 106 |
| 2018 | 99.09 | 0.91 | 110 |
| 2019 | 99.34 | 0.66 | 151 |
| 2020 | 98.35 | 1.65 | 121 |
| 2021 | 100.00 | 0.00 | 107 |
| Total | 98.54 | 1.46 | 1,093 |

Data originates from opentender.eu (https://opentender.eu/de) for the years 2011 to 2021, based on the Common Procurement Vocabulary (CPV) system for standardized classifications in public procurement, specifically category 60112000 (Public Transport – Road). The data is restricted to Germany, with duplicate entries counted only once. Only observations with a link to the call for tenders published on Tenders Electronic Daily (TED) are considered.

Table 3: Wage Mean Difference Test by Event Time

**West: Imputed Daily Wage**

|  | Treatment Group | | Control Group | | Difference |
|---|---|---|---|---|---|
|  | (1) | (2) | (3) | (4) | (2) – (4) |
| Event time | N | Mean | N | Mean |  |
| Before (t: -7 to -1) | 945 | 108.69 | 2219 | 105.35 | 3.34*** |
| After (t: 0 to 5) | 810 | 119.78 | 1,902 | 113.17 | 6.61*** |

**East: Imputed Daily Wage**

|  | Treatment Group | | Control Group | | Difference |
|---|---|---|---|---|---|
|  | (1) | (2) | (3) | (4) | (2) – (4) |
| Event time | N | Mean | N | Mean |  |
| Before (t: -7 to -1) | 756 | 90.54 | 1260 | 93.79 | -3.25*** |
| After (t: 0 to 5) | 648 | 97.48 | 1080 | 98.14 | -0.66 |

Note: The table presents the average daily wage, adjusted and deflated to 2015, for two time periods: five years preceding the law's introduction (t: -7 to -1) and five years following its implementation (t: 0 to 5). Figures are shown separately for the treatment and control groups in both West and East Germany. The mean t-test examines the differences between the two groups in the respective periods for statistical significance. Significance level: * p<0.10, ** p<0.05, *** p<0.01

Table 4: Wild Cluster Bootstrap inference by West and East Germany

| | West Germany | | | | | East Germany | | | | |
|---|---|---|---|---|---|---|---|---|---|---|
| | (1) | (2) | (3) | (4) | (5) | (6) | (7) | (8) | (9) | (10) |
| | | Bootstrap, by | | | | | Bootstrap, by | | | |
| | | federal state p-value | | individual p-value | | | federal state p-value | | individual p-value | |
| Event Time | Coefficient clustered federal state level | WCR | WCU | WCR | WCU | Coefficiet clustered federal state level | WCR | WCU | WCR | WCU |
| Ref. t-1 | | | | | | | | | | |
| ≤-6 | -0.0335*** (0.0040) | 0.066 | 0.000 | 0.015 | 0.016 | -0.0070 (0.0201) | 0.771 | 0.682 | 0.665 | 0.665 |
| -5 | -0.0479*** (0.0040) | 0.034 | 0.001 | 0.000 | 0.000 | 0.0014 (0.0151) | 0.905 | 0.900 | 0.923 | 0.923 |
| -4 | -0.0427*** (0.0052) | 0.048 | 0.000 | 0.000 | 0.000 | -0.0169 (0.0129) | 0.496 | 0.358 | 0.121 | 0.122 |
| -3 | -0.0378*** (0.0068) | 0.052 | 0.000 | 0.000 | 0.000 | -0.0143 (0.0204) | 0.590 | 0.503 | 0.201 | 0.201 |
| -2 | -0.0381** (0.0099) | 0.057 | 0.000 | 0.000 | 0.000 | -0.0036 (0.0035) | 0.399 | 0.273 | 0.625 | 0.625 |
| 0 | -0.0129 (0.0134) | 0.432 | 0.352 | 0.293 | 0.293 | 0.0203*** (0.0043) | 0.029 | 0.000 | 0.007 | 0.007 |
| 1 | -0.0231* (0.0087) | 0.077 | 0.000 | 0.027 | 0.028 | 0.0315** (0.0078) | 0.053 | 0.037 | 0.000 | 0.000 |
| 2 | -0.0070 (0.0124) | 0.766 | 0.685 | 0.568 | 0.567 | 0.0294** (0.0096) | 0.126 | 0.043 | 0.004 | 0.004 |
| 3 | -0.0075 (0.0081) | 0.418 | 0.342 | 0.601 | 0.601 | 0.0216** (0.0064) | 0.094 | 0.066 | 0.055 | 0.057 |
| 4 | -0.0027 (0.0113) | 0.866 | 0.829 | 0.865 | 0.865 | 0.0269** (0.0089) | 0.040 | 0.076 | 0.025 | 0.025 |
| 5 | 0.0079 (0.0137) | 0.601 | 0.547 | 0.606 | 0.607 | 0.0343** (0.0077) | 0.043 | 0.006 | 0.014 | 0.013 |
| R²adj | 0.915 | | | | | 0.909 | | | | |
| N | 5876 | | | | | 3744 | | | | |

Note: The table presents the results of a dynamic two-way fixed-effects event study, including personal and establishment fixed effects, on the log of imputed daily wages deflated to the year 2015. Columns 1 to 5 show results for West Germany, and columns 6 to 10 for East Germany. Columns 1 and 6 report the coefficients of the interaction term between event time (t−7 to t+5) and the treatment dummy (1 = federal state with collective bargaining compliance law, 0 = otherwise), using t−1 as the reference period. Clustered standard errors at the federal state level are reported in parentheses. Columns 2 to 5 and 7 to 10 display the p-values corresponding to the coefficients from columns 1 and 6, based on wild cluster bootstrap procedures with restricted heteroscedasticity (WCR) and unrestricted heteroscedasticity (WCU), clustered at the level indicated in the column headers. The bootstrap procedure uses Webb weights. * $p<0.10$, ** $p<0.05$, *** $p<0.01$

Figure 1: Distribution of Public Transport Employees by Law Introduction and Treatment Status, Separated for West and East Germany

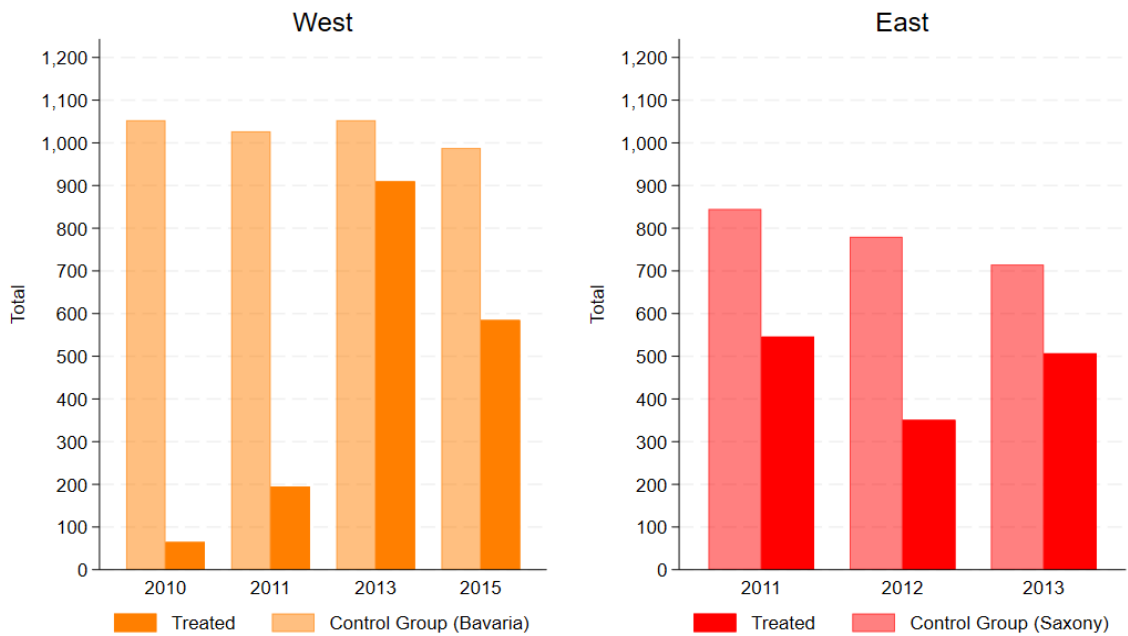

Note: The figure shows the distribution of observations by law introduction and enactment, separately for West and East Germany, according to treatment status. In West Germany, the laws were introduced in Saarland (2010), Rhineland-Palatinate (2011), Baden-Württemberg (2013), and Hesse (2015); Bavaria is the control group to the western federal states. In East Germany, the laws were introduced in Mecklenburg-Western Pomerania and Thuringia (2011), Brandenburg (2012), and Saxony-Anhalt (2012), Saxony is the control group to the eastern federal states.

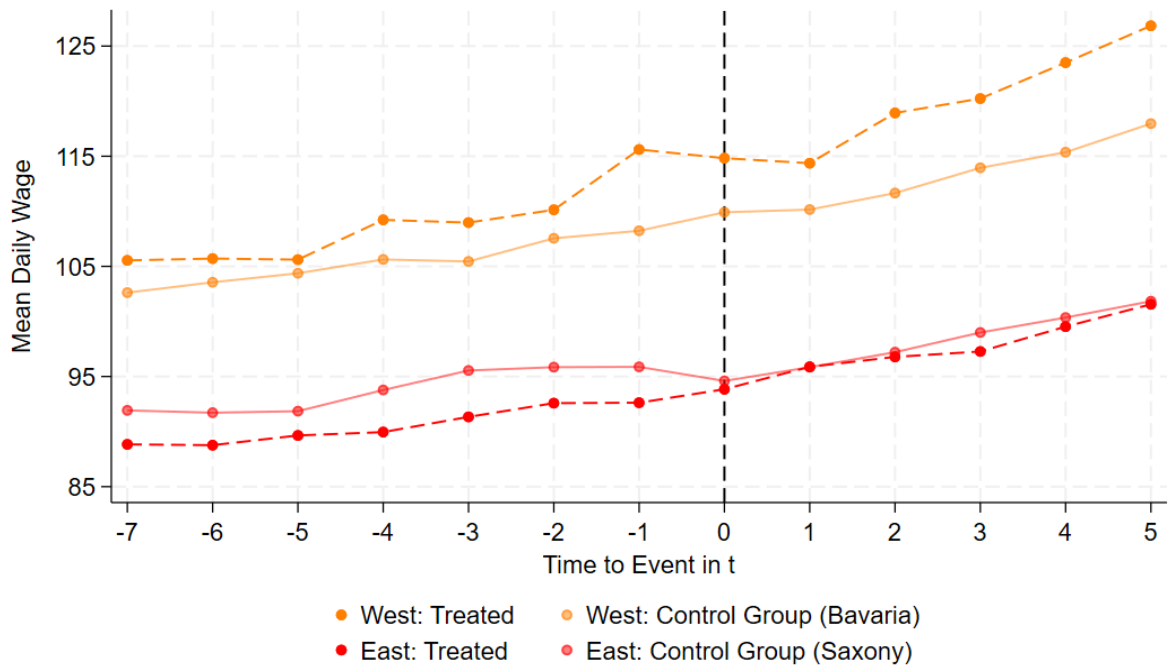

Figure 2: Wage Development by Treatment Status in West and East Germany

Note: The figure shows the average wage development from seven years (t-7) before to five years (t+5) after the implementation of collective bargaining compliance laws for full time employees in public transport, separately for West and East Germany. The dashed line represents federal states with collective bargaining compliance laws (Treated), while the solid line represents those that never introduced such laws (Never Treated). In West Germany, this applies to Bavaria, and in East Germany to Saxony.

Figure 3: Relationship Between Collective Bargaining Compliance Laws and Wages

West Germany

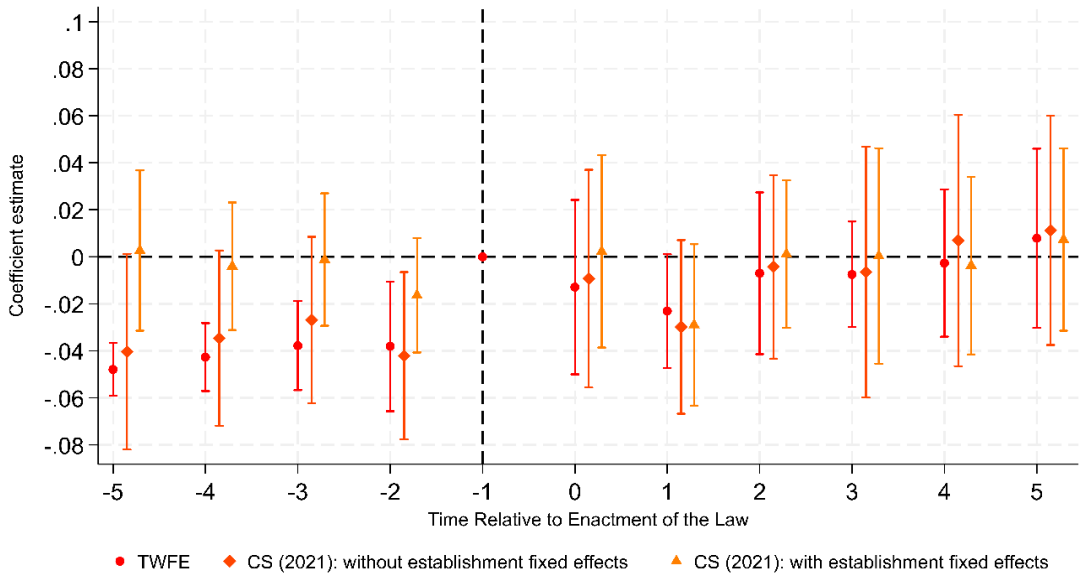

East Germany

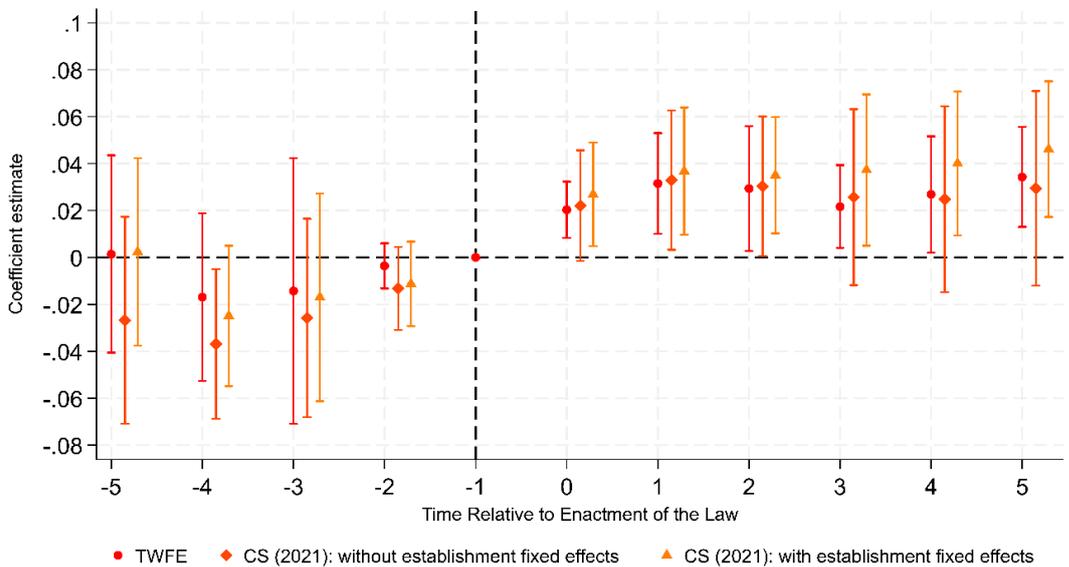

Note: The figure shows estimated coefficients for the interaction between the treatment indicator (1 = federal state with a collective bargaining compliance law, 0 = without) and the event period dummy (reference: t−1) for West and East Germany. The dependent variable is the log of imputed daily wages deflated to the year 2015. Two-way fixed effects (TWFE) estimates include year, person, and establishment fixed effects. Standard errors are clustered at the federal state level and show the 95% pointwise confidence intervals. Callaway and Sant'Anna (2021) (CS (2021)) estimates were conducted with and without establishment fixed effects. Confidence intervals for the Callaway and Sant'Anna (2021) estimates show 95% uniform confidence intervals based on a multiplicative wild bootstrap procedure. Sample sizes: TWFE: N-West: 5,876; N-East: 3,744; Callaway and Sant'Anna (2021): N-West: 2,352; N-East: 2,205.

Figure 4: Population Growth Index (2015 = 100) by West and East German Federal States

West Germany

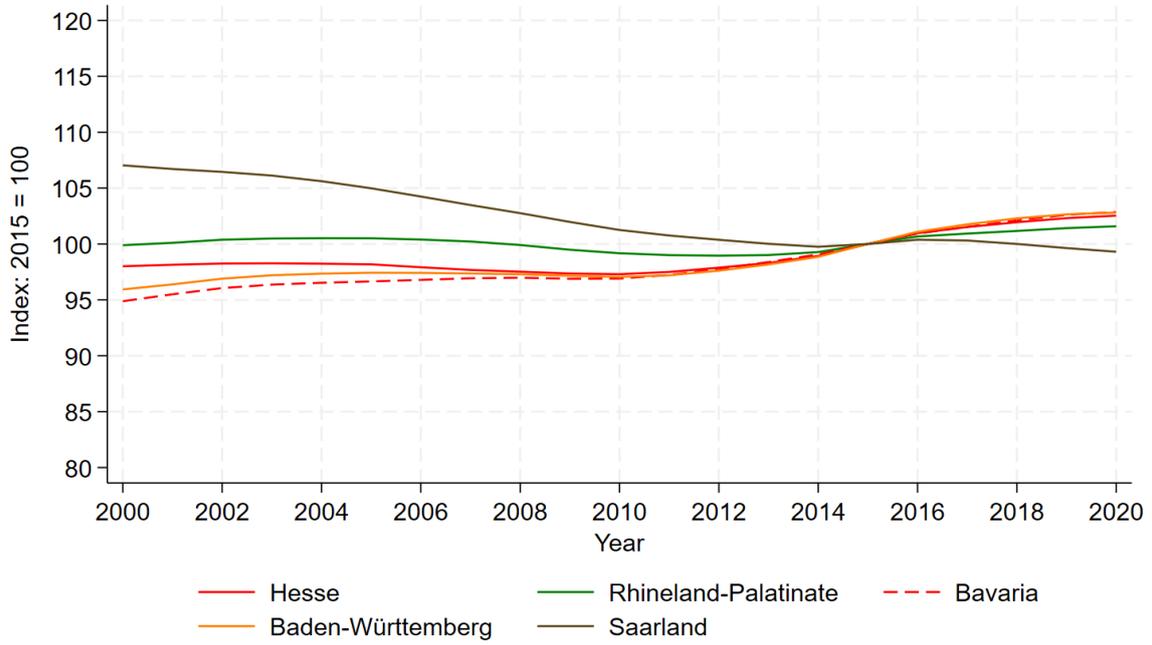

East Germany

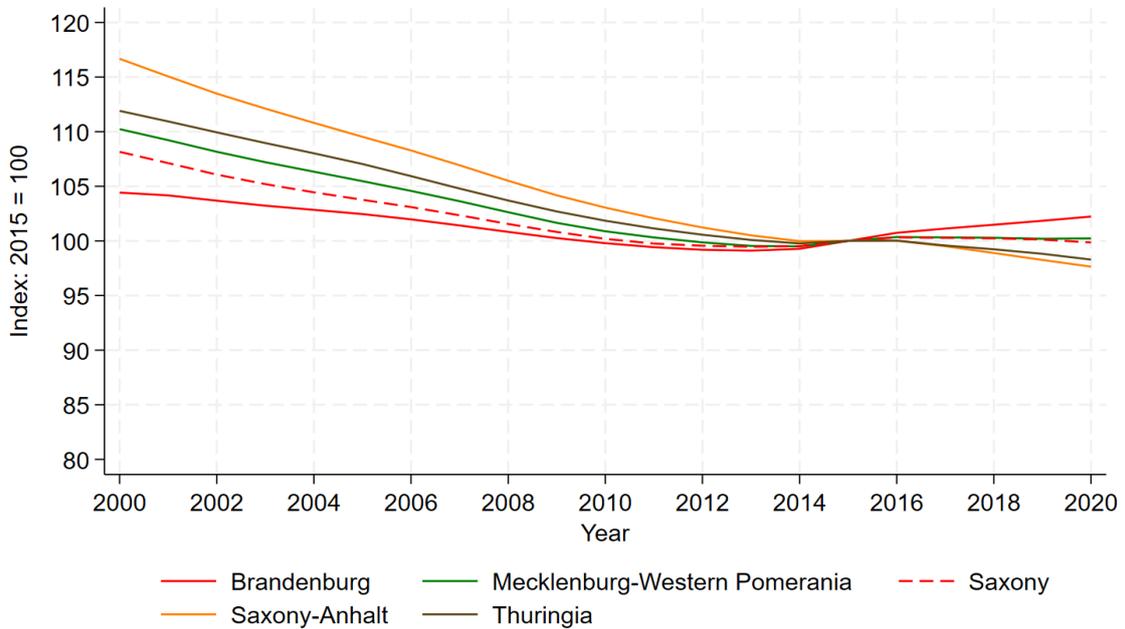

Note: The figure shows the population growth index relative to the population level in 2015 (2015 = 100) for West and East German federal states. The data is sourced from the National Accounts of the Federal States (Statistisches Landesamt Baden-Württemberg, 2023).

# Appendix

Figure A1: Relationship Between Collective Bargaining Compliance Laws and Wages Excluding Daily Wages Below EUR 68

West Germany

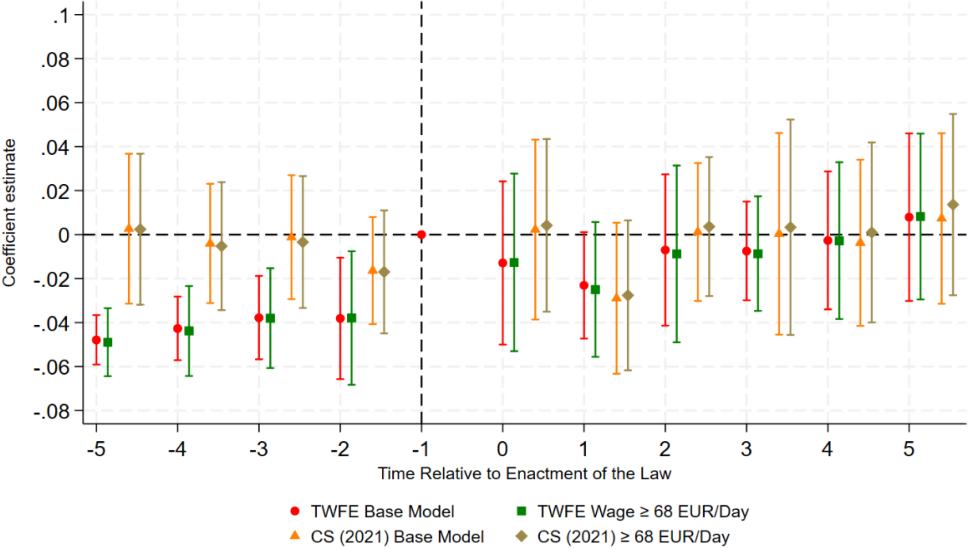

East Germany

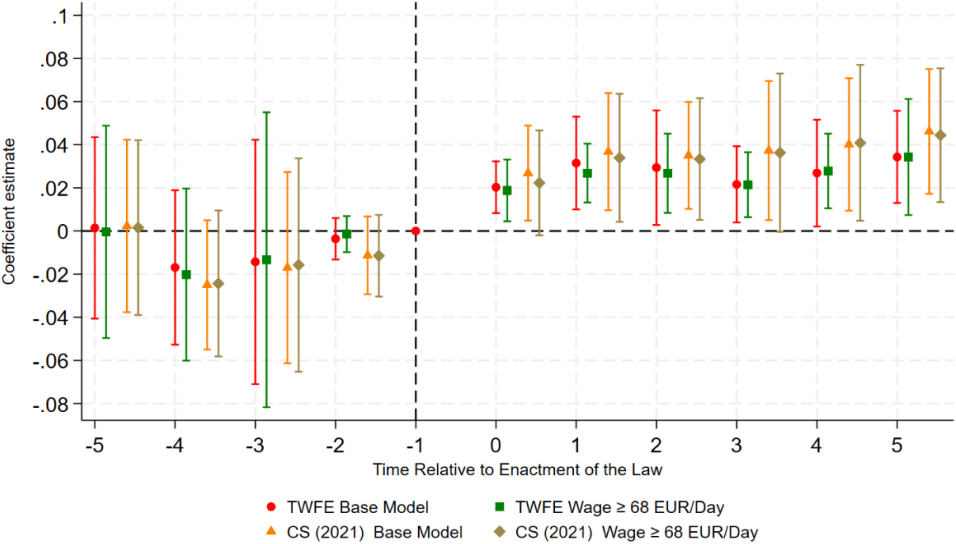

Note: The figure shows estimated coefficients for the interaction between the treatment indicator (1 = federal state with a collective bargaining compliance law, 0 = without) and the event period dummy (reference: t−1) for West and East Germany, considering both the main analysis and the robustness check excluding wages below 68 EUR per day. The dependent variable is the log of imputed daily wages deflated to the year 2015. Two-way fixed effects (TWFE) estimates include year, person, and establishment fixed effects. Standard errors are clustered at the federal state level and show the 95% pointwise confidence intervals. Callaway and Sant'Anna (2021) (CS (2021)) estimates include establishment fixed effects. Confidence intervals for the Callaway and Sant'Anna (2021) estimates show 95% uniform confidence intervals based on a multiplicative wild bootstrap procedure. Sample sizes: TWFE Base Model: N-West: 5,876; N-East: 3,744; TWFE Wage ≥ 68 EUR/Day: N-West: 5,655; N-East: 3,289; Callaway and Sant'Anna (2021) Base:N- West: 2,352; N-East: 2,205; Callaway and Sant'Anna (2021) Wage ≥ 68 EUR/Day: N-West: 2,268; N-East: 1,932.

Figure A2: West Germany: Examining Outliers in the Relationship Between Collective Bargaining Compliance Laws and Wages by Excluding Treated States

Two-way fixed effects (TWFE)

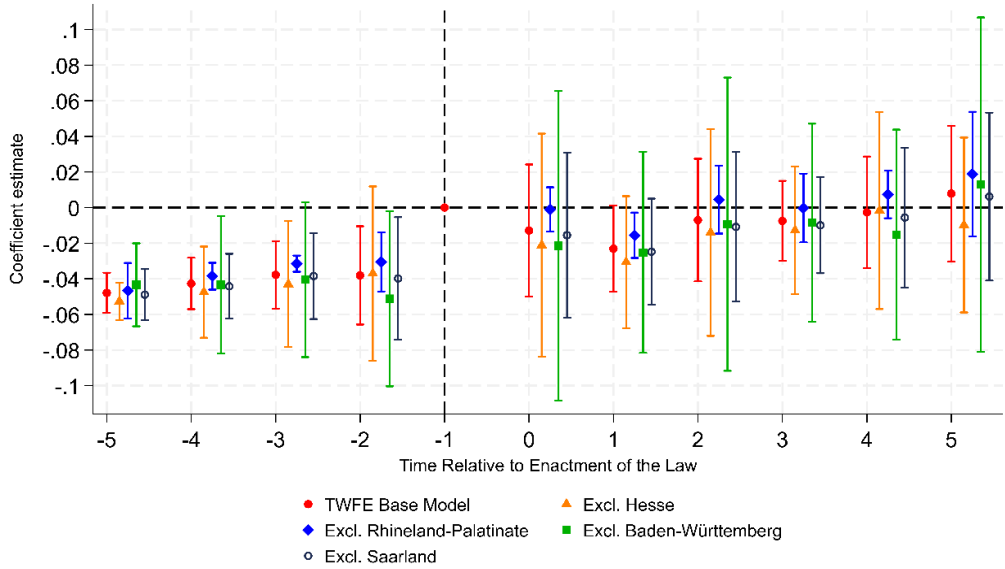

Callaway and Sant'Anna (2021)

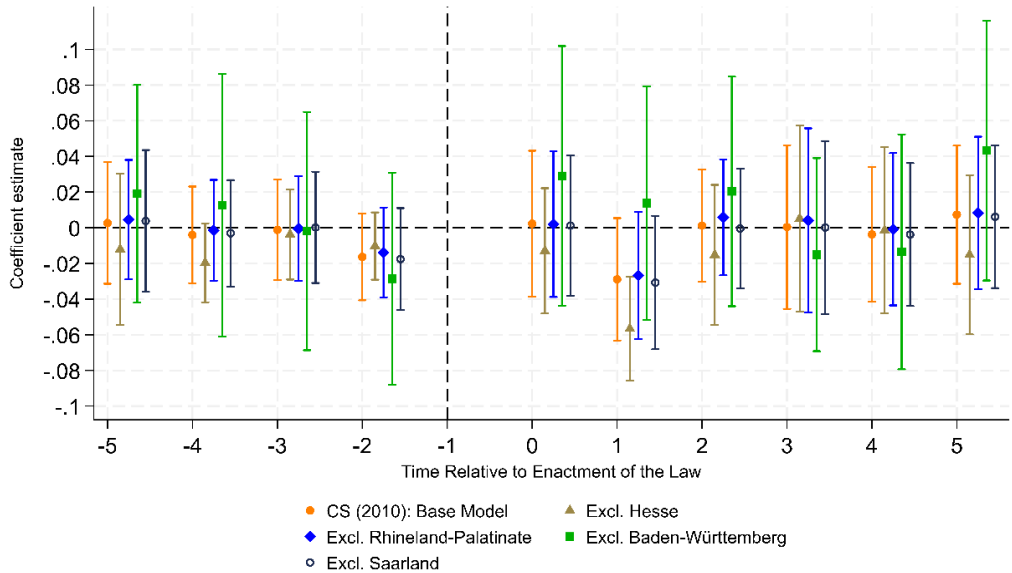

Note: The figure shows estimated coefficients for the interaction between the treatment indicator (1 = federal state with a collective bargaining compliance law, 0 = without) and the event period dummy (reference: t−1) for West Germany, considering both the main analysis and the robustness check excluding certain federal states. The dependent variable is the log of imputed daily wages deflated to the year 2015. Two-way fixed effects (TWFE) estimates include year, person, and establishment fixed effects. Standard errors are clustered at the federal state level and show the 95% pointwise confidence intervals. Callaway and Sant'Anna (2021) (CS (2021)) estimates also include establishment fixed effects. Confidence intervals for the Callaway and Sant'Anna (2021) estimates show 95% uniform confidence intervals based on a multiplicative wild bootstrap procedure. Sample sizes: TWFE estimations: N-Base Model: 5,876; N-Excl. Hessen: 5,291; N-Excl. Rhineland-Palatinate: 5,681; N-Excl. Baden-Württemberg: 4,966; N-Excl. Saarland: 5,811. Callaway and Sant'Anna (2021) estimations: N-Base Model: 2,352; N-Excl. Hessen: 1,890; N-Excl. Rhineland-Palatinate: 2,268; N-Excl. Baden-Württemberg: 1,554; N-Excl. Saarland: 2,310.

Figure A3: East Germany: Examining Outliers in the Relationship Between Collective Bargaining Compliance Laws and Wages by Excluding Treated States

Two-way fixed effects (TWFE)

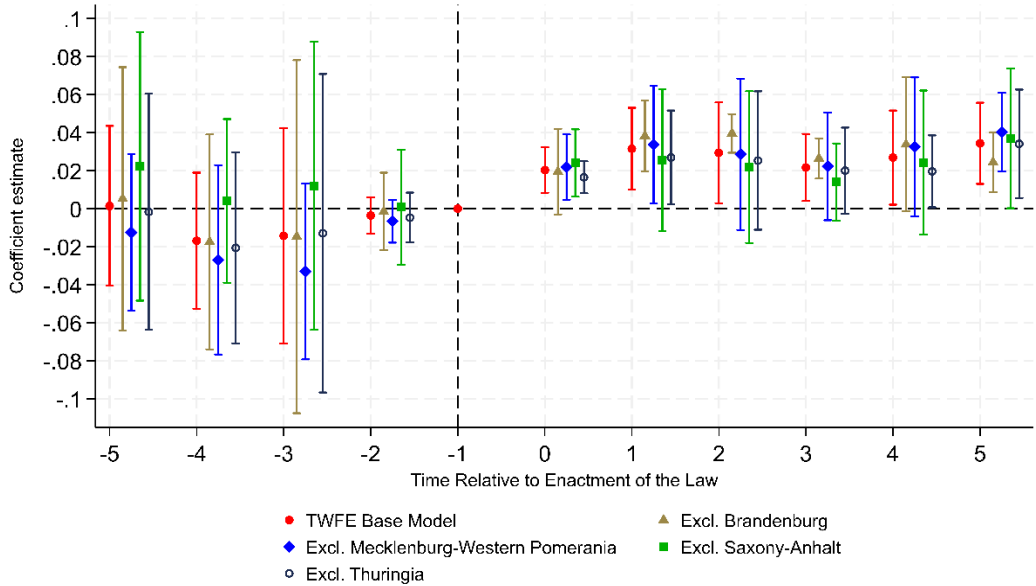

Callaway and Sant'Anna (2021)

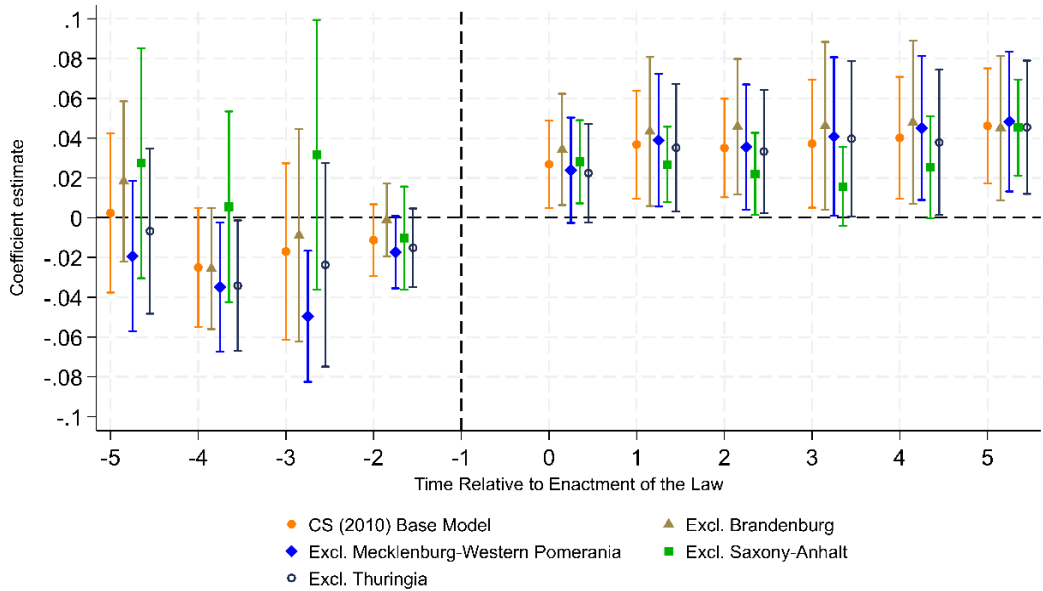

Note: The figure shows estimated coefficients for the interaction between the treatment indicator (1 = federal state with a collective bargaining compliance law, 0 = without) and the event period dummy (reference: t−1) for East Germany, considering both the main analysis and the robustness check excluding certain federal states. The dependent variable is the log of imputed daily wages deflated to the year 2015. Two-way fixed effects (TWFE) estimates include year, person, and establishment fixed effects. Standard errors are clustered at the federal state level and show the 95% pointwise confidence intervals. Callaway and Sant'Anna (2021) (CS (2021)) estimates also include establishment fixed effects. Confidence intervals for the Callaway and Sant'Anna (2021) estimates show 95% uniform confidence intervals based on a multiplicative wild bootstrap procedure. Sample sizes: TWFE estimations: N-Base Model: 3,744; N-Excl. Brandenburg: 3,393; N-Excl. Mecklenburg-Western Pomerania: 3,458; N-Excl. Saxony-Anhalt: 3,237; N-Excl. Thuringia: 3,484. Callaway and Sant'Anna (2021) estimations: N-Base Model: 2,205; N-Excl. Brandenburg: 1,890; N-Excl. Mecklenburg-Western Pomerania: 1,974; N-Excl. Saxony-Anhalt: 1,596; N-Excl. Thuringia: 2,016.

Figure A4: Testing the Relationship Between Collective Bargaining Compliance Laws and Wages Using a Placebo Event Year

West Germany

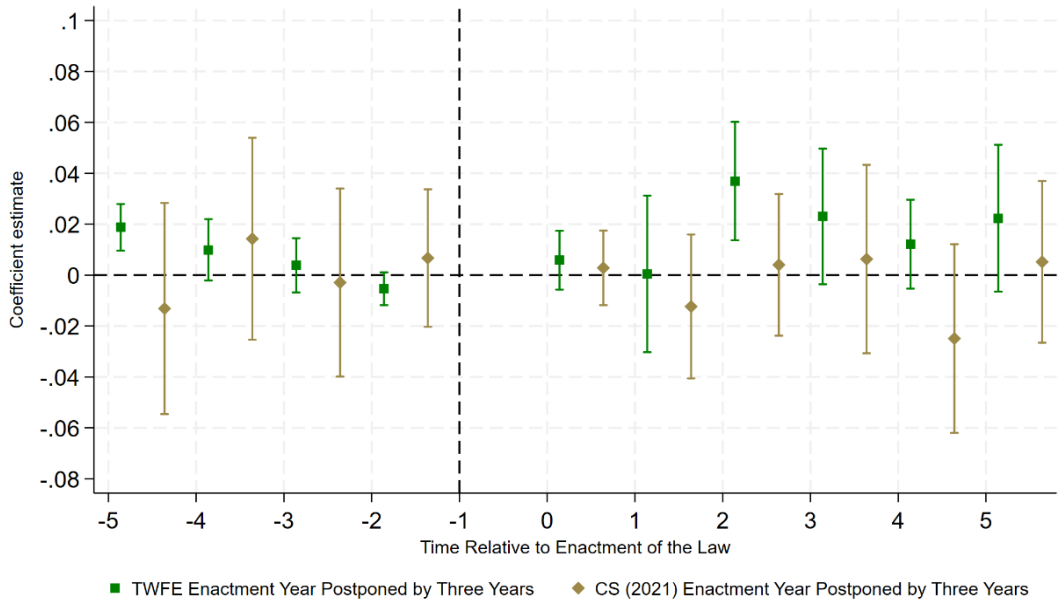

East Germany

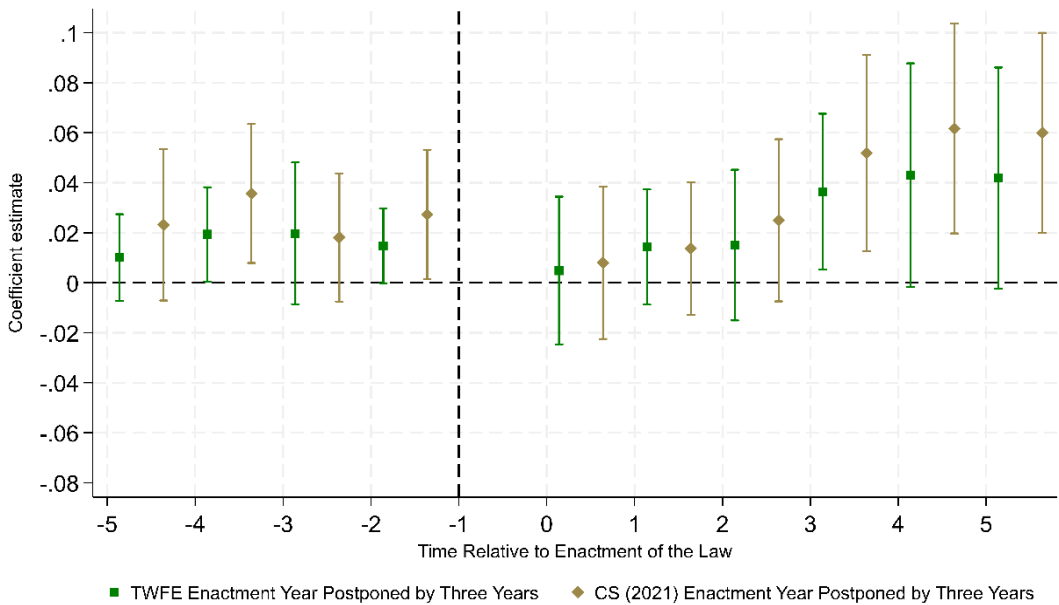

Note: The figure shows estimated coefficients for the interaction between the treatment indicator (1 = federal state with a collective bargaining compliance law; 0 = without) and the event period dummy (reference: t−1) for West and East Germany, considering the robustness check where the year of enactment is postponed by three years. The dependent variable is the log of imputed daily wages deflated to the year 2015. Two-way fixed effects (TWFE) estimates include year, person, and establishment fixed effects. Standard errors are clustered at the federal state level and show the 95% pointwise confidence intervals. Callaway and Sant'Anna (2021) (CS 2021) estimates include establishment fixed effects. Confidence intervals for the Callaway and Sant'Anna (2021) estimates show 95% uniform confidence intervals based on a multiplicative wild bootstrap procedure. Sample sizes: TWFE: N-West: 6,097; N-East: 4,316; Callaway and Sant'Anna (2021): N-West: 2,352; N-East: 2,205.

Online Appendix

| Table AO 1: | (1) | (2) |
|---|---|---|
| | West | East |
| Dependent variable: log daily imputed wage | Germany | Germany |
| Time Relative to Enactment (Reference t -1) | | |
| ≤ -6 | -0.0032 | 0.0082 |
| | (0.0167) | (0.0294) |
| -5 | -0.0001 | 0.0067 |
| | (0.0128) | (0.0262) |
| -4 | 0.0029 | 0.0093 |
| | (0.0084) | (0.0238) |
| -3 | 0.0006 | 0.0084 |
| | (0.0066) | (0.0195) |
| -2 | 0.0008 | 0.0044 |
| | (0.0027) | (0.0125) |
| 0 | -0.0028 | -0.0117 |
| | (0.0024) | (0.0166) |
| 1 | -0.0042 | -0.0230 |
| | (0.0057) | (0.0327) |
| 2 | -0.0065 | -0.0382 |
| | (0.0074) | (0.0499) |
| 3 | -0.0039 | -0.0539 |
| | (0.0117) | (0.0683) |
| 4 | -0.0061 | -0.0744 |
| | (0.0134) | (0.0880) |
| 5 | -0.0091 | -0.0893 |
| | (0.0135) | (0.1059) |
| Interaction term: Treatment dummy * Time Relative to Enactment (Reference t-1) | | |
| ≤ -6 | -0.0335*** | -0.0070 |
| | (0.0040) | (0.0201) |
| -5 | -0.0479*** | 0.0014 |
| | (0.0040) | (0.0151) |
| -4 | -0.0427*** | -0.0169 |
| | (0.0052) | (0.0129) |
| -3 | -0.0378*** | -0.0143 |
| | (0.0068) | (0.0204) |
| -2 | -0.0381** | -0.0036 |
| | (0.0099) | (0.0035) |
| 0 | -0.0129 | 0.0203*** |
| | (0.0134) | (0.0043) |
| 1 | -0.0231* | 0.0315** |
| | (0.0087) | (0.0078) |
| 2 | -0.0070 | 0.0294** |
| | (0.0124) | (0.0096) |
| 3 | -0.0075 | 0.0216** |
| | (0.0081) | (0.0064) |
| 4 | -0.0027 | 0.0269** |
| | (0.0113) | (0.0089) |
| 5 | 0.0079 | 0.0343** |
| | (0.0137) | (0.0077) |
| Constant | 4.6107*** | 4.5300*** |
| | (0.0235) | (0.0344) |
| Fixed Effect | | |
| Individual fixed effect | ✓ | ✓ |
| Year dummies | ✓ | ✓ |
| Establishment dummies | ✓ | ✓ |
| R-adj | 0.915 | 0.909 |
| N | 5876 | 3744 |

Note: The table shows the results of a dynamic two-way fixed effects event study. The dependent variable is the log of imputed daily wages deflated to the year 2015. Standard errors are clustered at federal state level.  * p<0.10, ** p<0.05, *** p<0.01

# References

Abadie, A., Athey, S., Imbens, G. W., & Wooldridge, J. M. (2022). When Should You Adjust Standard Errors for Clustering? *The Quarterly Journal of Economics*, *138*(1), 1–35. https://doi.org/10.1093/qje/qjac038

Addison, J., Teixeira, P., Evers, K., & Bellmann, L. (2014). Indicative and Updated Estimates of the Collective Bargaining Premium in Germany. *Industrial Relations: A Journal of Economy and Society*, *53*(1), 125–156. https://doi.org/10.1111/irel.12049

Baumgarten, D., Felbermayr, G., & Lehwald, S. (2020). Dissecting Between-Plant and Within-Plant Wage Dispersion: Evidence from Germany. *Industrial Relations: A Journal of Economy and Society*, *59*(1), 85–122. https://doi.org/10.1111/irel.12249

Beeson, M., Wildman, J. M., & Wildman, J. (2024). Does tackling poverty related barriers to education improve school outcomes? Evidence from the North East of England. *Economics Letters*, *236*, 111614. https://doi.org/10.1016/j.econlet.2024.111614

Bonaccolto-Töpfer, M., & Schnabel, C. (2023). Is There a Union Wage Premium in Germany and Which Workers Benefit Most? *Economies*, *11*(2), 50. https://www.mdpi.com/2227-7099/11/2/50

Borusyak, K., Jaravel, X., & Spiess, J. (2024). Revisiting Event-Study Designs: Robust and Efficient Estimation. *The Review of Economic Studies*, *91*(6), 3253–3285. https://doi.org/10.1093/restud/rdae007

Bossler, M., & Gerner, H.-D. (2019). Employment Effects of the New German Minimum Wage: Evidence from Establishment-Level Microdata. *ILR Review*, *73*(5), 1070–1094. https://doi.org/10.1177/0019793919889635

Bossler, M., & Schank, T. (2023). Wage Inequality in Germany after the Minimum Wage Introduction. *Journal of Labor Economics*, *41*(3), 813–857. https://doi.org/10.1086/720391

Brandt, T., & Schulten, T. (2008). Auswirkungen von Privatisierung und Liberalisierung auf die Tarifpolitik in Deutschland. In B. Torsten, S. Thorsten, S. Gabriele, & W. Jörg (Eds.), *Europa im Ausverkauf - Liberalisierung und Privatisierung öffentlicher Dienstleistungen und ihre Folgen für die Tarifpolitik* (pp. 68 – 91).

Bundesministerium für Arbeit und Soziales. (2023). *Öffentliche Konsultation zur Bindung der Auftragsvergabe des Bundes an die Einhaltung von Tarifverträgen ("Bundes-Tariftreue") durch das Bundesministerium für Wirtschaft und Klimaschutz und das Bundesministerium für Arbeit und Soziales.* https://www.bmas.de/DE/Arbeit/Arbeitsrecht/Staerkung-der-Tarifbindung/Oeffentliche-Konsultation-zur-Tariftreue/oeffentliche-konsultation-zur-tariftreue.html

Bundesministerium für Arbeit und Soziales. (2024). *Entwurf eines Gesetzes zur Stärkung der Tarifautonomie durch die Sicherung von Tariftreue bei der Vergabe öffentlicher Aufträge des Bundes (Tariftreuegesetz)*. https://www.bmas.de/SharedDocs/Downloads/DE/Gesetze/Regierungsentwuerfe/reg-tariftreuegesetz.pdf?__blob=publicationFile&v=1

Bundesverfassungsgericht. (2006). *Beschluss vom 1. Juli 2006, 1BvL 4/00*. https://www.bundesverfassungsgericht.de/entscheidungen/ls20060711_1bvl000400.html

Burauel, P., Caliendo, M., Grabka, M. M., Obst, C., Preuss, M., Schröder, C., & Shupe, C. (2020). The Impact of the German Minimum Wage on Individual Wages and Monthly Earnings. *Jahrbücher für Nationalökonomie und Statistik*, *240*(2-3), 201–231. https://doi.org/doi:10.1515/jbnst-2018-0077

Callaway, B., & Sant'Anna, P. H. C. (2021). Difference-in-Differences with multiple time periods. *Journal of Econometrics*, *225*(2), 200–230. https://doi.org/10.1016/j.jeconom.2020.12.001

Cameron, A. C., Gelbach, J. B., & Miller, D. L. (2008). Bootstrap-Based Improvements for Inference with Clustered Errors. *The Review of Economics and Statistics*, *90*(3), 414–427. https://doi.org/10.1162/rest.90.3.414

Card, D., & Krueger, A. B. (2000). Minimum Wages and Employment: A Case Study of the Fast-Food Industry in New Jersey and Pennsylvania: Reply. *The American Economic Review*, *90*(5), 1397–1420. http://www.jstor.org/stable/2677856

Caspers, G. (2024). Tarifanwendung bei der öffentlichen Auftragsvergabe. *Zeitschrift für Arbeitsrecht*, *55*(2), 225–244. https://doi.org/doi:10.9785/zfa-2024-550205

CDU, CSU, & SPD. (2025). *Verantwortung für Deutschland. Koalitionsvertrag zwischen CDU, CSU und SPD für die 21. Legislaturperiode*. https://www.spd.de/fileadmin/Dokumente/Koalitionsvertrag_2025.pdf

Datta, N., & Machin, S. (2024). Government Contracting and Living Wages > Minimum Wages. *IZA Discussion Paper No. 17117*.

de Chaisemartin, C., & D'Haultfœuille, X. (2020). Two-Way Fixed Effects Estimators with Heterogeneous Treatment Effects. *American Economic Review*, *110*(9), 2964–2996. https://doi.org/10.1257/aer.20181169

de Chaisemartin, C., & D'Haultfœuille, X. (2023). Two-way fixed effects and differences-in-differences with heterogeneous treatment effects: a survey. *The Econometrics Journal*, *26*(3), C1–C30. https://doi.org/10.1093/ectj/utac017

Djogbenou, A., MacKinnon, J., & Nielsen, M. (2019). Asymptotic theory and wild bootstrap inference with clustered errors. *Journal of Econometrics*, *212*(2), 393–412. https://EconPapers.repec.org/RePEc:eee:econom:v:212:y:2019:i:2:p:393-412

Drechsler, J., & Ludsteck, J. (2024). *Bridging Between Different BeH Industry Classifications via Imputation*. https://ideas.repec.org/p/iab/iabfme/202404(en).html

Dustmann, C., Lindner, A., Schönberg, U., Umkehrer, M., & vom Berge, P. (2021). Reallocation Effects of the Minimum Wage. *The Quarterly Journal of Economics*, *137*(1), 267–328. https://doi.org/10.1093/qje/qjab028

European Court of Auditors. (2023). *Special report 28/2023: Public procurement in the EU – Less competition for contracts awarded for works, goods and services in the 10 years up to 2021*. https://www.eca.europa.eu/en/publications/sr-2023-28&od=1

European Union. (2007). *Regulation (EC) No 1370/2007 of the European Parliament and of the Council of 23 October 2007 on public passenger transport services by rail and by road and repealing Council Regulations (EEC) Nos 1191/69 and 1107/70*. https://eur-lex.europa.eu/legal-content/EN/TXT/?uri=CELEX%3A32007R1370

European Union. (2016). *Consolidated versions of the Treaty on European Union and the Treaty on the Functioning of the European Union Consolidated version of the Treaty on European Union Consolidated version of the Treaty on the Functioning of the European Union Protocols Annexes to the Treaty on the Functioning of the European Union Declarations annexed to the Final Act of the Intergovernmental Conference which adopted the Treaty of Lisbon, signed on 13 December 2007 Tables of equivalences*. https://eur-lex.europa.eu/legal-content/EN/TXT/?uri=celex%3A12016ME%2FTXT

Goodman-Bacon, A. (2021). Difference-in-differences with variation in treatment timing. *Journal of Econometrics*, *225*(2), 254–277. https://doi.org/https://doi.org/10.1016/j.jeconom.2021.03.014

Grüttner, A., Lenk, T., & Rottmann, O. (2012). Aspekte zur Konzipierung von vergleichenden Effizienzberechnungen verschiedener Vergabeverfahren im ÖPNV. *Zeitschrift für öffentliche und gemeinwirtschaftliche Unternehmen: ZögU / Journal for Public and Nonprofit Services*, *35*(1), 28–43. http://www.jstor.org/stable/41634748

Hirsch, B., & Mueller, S. (2020). Firm Wage Premia, Industrial Relations, and Rent Sharing in Germany. *ILR Review*, *73*(5), 1119–1146. https://doi.org/10.1177/0019793920917105

Hohendanner, C., & Kohaut, S. (2024). *75 Jahre Tarifvertragsgesetz: Sind Branchentarife und betriebliche Mitbestimmung ein Auslaufmodell?* . IAB-Forum 22. April 2024. Retrieved 13. March 2025 from https://www.iab-forum.de/75-jahre-tarifvertragsgesetz-sind-branchentarife-und-betriebliche-mitbestimmung-ein-auslaufmodell/

Jäger, S., Naidu, S., & Schoefer, B. (2024). Collective Bargaining, Unions, and the Wage Structure: An International Perspective. *National Bureau of Economic Research Working Paper Series*, *No. 33267*. http://www.nber.org/papers/w33267

Johnson, M. (2017). Implementing the living wage in UK local government. *Employee Relations*, *39*, 840–849. https://doi.org/10.1108/ER-02-2017-0039

MacKinnon, J. G., & Webb, M. D. (2017). Wild Bootstrap Inference for Wildly Different Cluster Sizes. *Journal of Applied Econometrics*, *32*(2), 233–254. https://doi.org/10.1002/jae.2508

MacKinnon, J. G., & Webb, M. D. (2018). The wild bootstrap for few (treated) clusters. *The Econometrics Journal*, *21*(2), 114–135. https://doi.org/10.1111/ectj.12107

McCrudden, C. (2011). The Rüffert Case and Public Procurement. In M. Cremona (Ed.), *Market Integration and Public Services in the European Union* (pp. 0). Oxford University Press. https://doi.org/10.1093/acprof:oso/9780199607730.003.0005

Miller, D. L. (2023). An Introductory Guide to Event Study Models. *Journal of Economic Perspectives*, *37*(2), 203–230. https://doi.org/10.1257/jep.37.2.203

Nassibi, G., Rödl, F., & Schulten, T. (2016). *Perspektiven vergabespezifischer Mindestlöhne nach dem Regio-Post-Urteil des EuGH*. https://ideas.repec.org/p/zbw/wsipbs/3.html

Neumark, D., Thompson, M., & Koyle, L. (2012). The effects of living wage laws on low-wage workers and low-income families: What do we know now? *IZA Journal of Labor Policy*, *1*(1), 11. https://doi.org/10.1186/2193-9004-1-11

Oberfichtner, M., & Schnabel, C. (2019). The German Model of Industrial Relations: (Where) Does It Still Exist? *Jahrbücher für Nationalökonomie und Statistik*, *239*(1), 5–37. https://doi.org/doi:10.1515/jbnst-2018-0158

Panahian Fard, D., Schmucker, A., Seth, S., Umkehrer, M., & Zimmermann, F. (2024). Linked-Employer-Employee-Data of the IAB: LIAB Longitudinal Model (LIAB LM) 1975-2021. *FDZ-Datenreport*, *04/2024*. https://doi.org/10.5164/IAB.FDZD.2404.en.v1

Peña, G., & Sanso-Navarro, M. (2025). The long-run population effects of floods in Spanish municipalities, 1877–2011. *Applied Economics*, 1–17. https://doi.org/10.1080/00036846.2024.2448611

Pham, L. D., Corcoran, S. P., Henry, G. T., & Zimmer, R. (2025). Do the Effects Persist? An Examination of Long-Term Effects After Students Leave Turnaround Schools. *American Educational Research Journal*, *62*(1), 180–213. https://doi.org/10.3102/00028312241284026

Resch, H. (2015). Branchenanalyse: Zukunft des ÖPNV. *Study der Hans-Böckler-Stiftung*, *302*.

Riphahn, R. T., & Schnitzlein, D. D. (2016). Wage mobility in East and West Germany. *Labour Economics*, *39*, 11–34. https://doi.org/10.1016/j.labeco.2016.01.003

Robertson, R. (2023). Labor compliance programs in developing countries and trade flows: Evidence from Better Work. *Economics Letters*, *228*, 111162. https://doi.org/10.1016/j.econlet.2023.111162

Roodman, D., Nielsen, M. Ø., MacKinnon, J. G., & Webb, M. D. (2019). Fast and wild: Bootstrap inference in Stata using boottest. *The Stata Journal*, *19*(1), 4–60. https://doi.org/10.1177/1536867X19830877

Roth, J., Sant'Anna, P. H. C., Bilinski, A., & Poe, J. (2023). What's trending in difference-in-differences? A synthesis of the recent econometrics literature. *Journal of Econometrics*, *235*(2), 2218–2244. https://doi.org/10.1016/j.jeconom.2023.03.008

Sack, D., & Sarter, E. (2018). Collective bargaining, minimum wages and public procurement in Germany: Regulatory adjustments to the neoliberal drift of a coordinated market economy. *Journal of Industrial Relations*, *60*(5), 669–690. https://doi.org/10.1177/0022185618795706

Schmucker, A., Seth, S., & Vom Berge, P. (2023). Sample of Integrated Labour Market Biographies (SIAB) 1975-2021. *FDZ-Datenreport*, *02/2023*. https://doi.org/10.5164/IAB.FDZD.2302.de.

Schnabel, C. (2016). United, Yet Apart? A Note on Persistent Labour Market Differences between Western and Eastern Germany. *Jahrbücher für Nationalökonomie und Statistik*, *236*(2), 157–179. https://doi.org/doi:10.1515/jbnst-2015-1012

Schulten, T. (2021). *Social clauses in German public procurement–towards a Post-Rüffert regime?* ETUC.

Statistisches Bundesamt (Destatis). (2025). *Vergabestatistik*. Retrieved 18.02.2025 from https://www.destatis.de/DE/Themen/Staat/Oeffentliche-Finanzen/Vergabestatistik/_inhalt.html#642852

Statistisches Landesamt Baden-Württemberg. (2023). Entstehung, Verteilung und Verwendung des Bruttoinlandsprodukts in den Ländern der Bundesrepublik Deutschland 1991 bis 2022 Länderergebnisse Band 5, Reihe 1 ; Berechnungsstand: August 2022/Februar 2023

Stüber, H., Dauth, W., & Eppelsheimer, J. (2023). A guide to preparing the sample of integrated labour market biographies (SIAB, version 7519 v1) for scientific analysis. *Journal for Labour Market Research*, *57*(1), 7. https://doi.org/10.1186/s12651-023-00335-w

Sun, L., & Abraham, S. (2021). Estimating dynamic treatment effects in event studies with heterogeneous treatment effects. *Journal of Econometrics*, *225*(2), 175–199. https://doi.org/10.1016/j.jeconom.2020.09.006

Thüsing, G. (2022). *Sozialstandards im Mobilitätsgewerbe* (1. ed.). Nomos.

Wooldridge, J. M. (2021). Two-Way Fixed Effects, the Two-Way Mundlak Regression, and Difference-in-Differences Estimators. https://doi.org/10.2139/ssrn.3906345

**In der Diskussionspapierreihe sind kürzlich erschienen:**

**Recently published Discussion Papers:**

Eine aktualisierte Liste der Diskussionspapiere findet sich auf der Homepage:
http://www.arbeitsmarkt.rw.fau.de/

An updated list of discussion papers can be found at the homepage:
http://www.arbeitsmarkt.rw.fau.de